\documentclass[12pt]{article}
\usepackage{amsmath}
\usepackage{amssymb}
\begin{document}
\vskip 2cm
\begin{center}
{\sf {\Large  Supersymmetrization of horizontality condition: nilpotent symmetries for a free spinning 
relativistic particle}}

\vskip 3.0cm

{\sf A. Shukla$^{(a)}$, S. Krishna$^{(a)}$, R. P. Malik$^{(a,b)}$}\\
$^{(a)}$ {\it Physics Department, Centre of Advanced Studies,}\\
{\it Banaras Hindu University, Varanasi - 221 005, (U.P.), India}\\

\vskip 0.1cm

{\bf and}\\

\vskip 0.1cm

$^{(b)}$ {\it DST Centre for Interdisciplinary Mathematical Sciences,}\\
{\it Faculty of Science, Banaras Hindu University, Varanasi - 221 005, India}\\
{\small {\sf {e-mails: ashukla038@gmail.com; skrishna.bhu@gmail.com;  malik@bhu.ac.in}}}

\end{center}

\vskip 2cm

\noindent
{\bf Abstract:} 
We clearly and consistently supersymmetrize the celebrated horizontality condition to 
derive the off-shell nilpotent and absolutely anticommuting Becchi-Rouet-Stora-Tyutin (BRST) and anti-BRST
symmetry transformations for the supersymmetric system of a free spinning relativistic particle within 
the framework of superfield approach to BRST formalism. For the precise determination of the 
proper (anti-)BRST symmetry transformations for all the bosonic and fermionic dynamical variables of our system,
we consider the present theory on a (1, 2)-dimensional supermanifold 
parameterized by an even (bosonic) variable ($\tau$) 
and a pair of odd (fermionic) variables $\theta$ and $\bar\theta$ (with $\theta^2 = \bar\theta^2 = 0,\; \theta \bar\theta + \bar\theta \theta = 0$) of the Grassmann algebra. One of the most important 
and novel features of our present investigation
is the derivation of  (anti-)BRST invariant Curci-Ferrari type restriction which turns out to be responsible 
for the absolute anticommutativity of the (anti-)BRST transformations and existence of the coupled (but equivalent) Lagrangians for the present theory of a supersymmetric system. These observations are completely
new results for this model.\\


\newpage

\noindent
\section {Introduction}

One of the most intuitive geometrical\footnote{To be precise, the geometrical approaches
adopted in [1-3] do not utilize superfields. Rather, these exploit the ordinary fields which
depend on the vertical directions of the fiber bundle. At their very best, these endeavors 
can be treated as a set of precursors to the superfield formalism proposed in [4-7].} 
approaches [1-3] to Becchi-Rouet-Stora-Tyutin (BRST) formalism
is the superfield formulation (see, e.g. [4-10]) where the abstract mathematical properties,
associated with the (anti-)BRST symmetries $s_{(a)b}$ of any arbitrary physical system,
find their geometrical origin and interpretation in the language of specific entities
of the superfield formalism. Within the framework of the latter formalism (i.e. superfield
formalism), a given $D$-dimensional physical theory (e.g. gauge theory, reparametrization
invariant theory, etc.) is considered on a $(D, 2)$-dimensional supermanifold parametrized
by the superspace variables $Z^M = (x^\mu, \theta, \bar\theta)$ where $x^\mu$ (with
$\mu = 0, 1, 2,...,D-1$) are the  ordinary bosonic coordinates of the ordinary flat
Minkowaskian spacetime  and $(\theta, \bar\theta)$ are a pair of Grassmannian
variables (with $\theta^2 = \bar \theta^2 = 0, \;\theta \bar \theta + \bar \theta \theta = 0$).
It has been well-established that the proper (i.e. nilpotent and absolutely anticommuting)
(anti-)BRST symmetry transformations of the
$D$-dimensional ordinary theory are connected with the translational generators along the
Grassmannian directions of the $(D, 2)$-dimensional supermanifold. These connections
have been established by exploiting the theoretical potential and power of the
so-called horizontality condition (HC).

In the context of $D$-dimensional $p$-form ($p = 1, 2, 3,...$) (non-)Abelian gauge theories, the super curvature 
$(p + 1)$-form [defined on the $(D, 2)$-dimensional supermanifold] is covariantly reduced 
to its counterpart--the ordinary curvature $(p + 1)$-form [defined on the ordinary $D$-dimensional
Minkowskian spacetime]. Physically, this reduction amounts to the fact that the gauge covariant quantity
(e.g. curvature term) remains independent of the Grassmannian variables of the $(D, 2)$-dimensional
supermanifold, on which, the ordinary $D$-dimensional (non-)Abelian $p$-form gauge theory is considered.
In the process of the above covariant reduction, the proper (i.e. off-shell nilpotent and
absolutely anticommuting) (anti-)BRST symmetries of the above $D$-dimensional theories emerge and they
are geometrically identified with the translational generators ($\partial_\theta, \partial_{\bar\theta}$)
along the Grassmannian
directions of the $(D, 2)$-dimensional supermanifold [4,5]. Exactly similar is the situation
with the reparametrization invariant theories where the analogue of the curvature term
is found and the above covariant reduction procedure generates the proper (anti-)BRST symmetry
transformations for the reparametrization invariant theory as well [9,10].

So far, the above superfield formalism has 
{\it not} been applied to the supersymmetric systems of some physical interest.   
As a consequence, to the best of our knowledge, one is not clear about the systematic generalization of
the HC for the description of a supersymmetric physical system within the framework of superfield formalism.
One of the main motivations of our present endeavor is to theoretically state the supersymmetric version 
of HC and apply it to the derivation of proper (anti-)BRST symmetry transformations for the supersymmetric
system of a one (0 + 1)-dimensional (1D) model of a free relativistic spinning particle which
represents an interesting toy model for the supergravity theories. In our present investigation, 
we derive the proper (anti-)BRST
symmetry transformations $s_{(a)b}$ for  the above supersymmetric system 
within the framework of augmented version of Bonora-Tonin (BT) superfield formalism\footnote{The BT
superfield formalism, applied to a given (non-)Abelian gauge theory, utilizes {\it only} the HC. 
In a set of papers (see, e.g. [8-10]), we have generalized the HC in a consistent
fashion by incorporating some other appropriate restrictions for the derivation of the
(anti-)BRST symmetry transformations for the gauge as well as 
matter fields of a given gauge/reparametrization invariant theory.}
where the HC, (super)gauge invariant restrictions [(S)GIR] and supersymmetric version  of HC (SUSY-HC)
are all exploited together. In fact, we are theoretically compelled to exploit all these
restrictions for the derivation of the full set of {\it proper} and precise (anti-)BRST symmetry transformations.

The off-shell nilpotency ($s^2_{(a)b} = 0$) and absolute anticommutativity property ($s_b\;s_{ab} + s_{ab}\; s_b = 0$)
of the (anti-)BRST transformations $s_{(a)b}$ are very sacrosanct because they physically imply 
the fermionic nature of $s_{(a)b}$ and the linear independence of $s_b$ and $s_{ab}$, respectively. 
The judicious combination of the (S)GIR and (SUSY-)HC leads to the derivation of such type of
proper (anti-)BRST symmetry transformations for the supersymmetric system of spinning relativistic particle. 
We have shown that the SUSY-HC [cf. (22)] resembles very much like the Maurer-Cartan equation (which defines 
the curvature tensor for the non-Abelian gauge theory). The other novel feature of our present investigation is the
derivation of (anti-)BRST invariant Curci-Ferrari (CF)-type restriction [cf. (30)] 
from the SUSY-HC of our present theory. It will be noted that the idea of (super)gauge 
invariant restrictions [(S)GIR] also 
plays a pivotal role in the derivation
of {\it all} the (anti-)BRST transformations for {\it all} the dynamical variables of our present
supersymmetric system.

In our present endeavor, we have captured all the key properties of (anti-)BRST symmetry transformations in the language of
specific objects of our augmented version of BT superfield formalism. For instance, the nilpotency and absolute anticommutativity
properties of the (anti-)BRST symmetry transformations as well as the (anti-)BRST invariance of the coupled (but equivalent) Lagrangians
have been incorporated within the framework of our superfield formalism. Finally, we have shown the (anti-)BRST invariance of the 
CF-type restriction (cf. Sec. 5) which plays a decisive role in our present investigation. In all these superfield descriptions, the
mappings between the (anti-)BRST symmetry transformations $s_{(a)b}$ and the translation generators ($\partial_\theta, \partial_{\bar\theta}$)
along the Grassmannian directions of the supermanifold
[cf. (43) below] have played a role of paramount importance. In fact, the nilpotency
and absolute anticommutativity properties of the BRST symmetries are automatically captured within our superfield
formalism because $\partial^2_{\theta} = \partial^2_{\bar\theta} = 0$, $\partial_{\theta}\;\partial_{\bar\theta} 
+ \partial_{\bar\theta}\;\partial_{\theta}  = 0$.

The main motivating factors  behind our present investigation are as follows. First, it is, 
for the first-time, that the augmented version of BT superfield formalism [8-10] is being 
applied to a supersymmetric model of a relativistic particle. Second, the generalization of HC
to the supersymmetric gauge theory (christened as SUSY-HC in our present work) was a challenging
problem which we have resolved in our present endeavor. In fact, we have obtained a neat expression for 
the SUSY-HC which resembles very much like the Maurer-Cartan equation [cf. (22)]. Third, a completely new result is the
derivation of CF-type restriction for the present supersymmetric system of spinning relativistic particle.
Finally, the derivation of the proper (i.e. off-shell nilpotent and absolutely anticommuting) 
anti-BRST symmetry transformations and coupled (but equivalent) Lagrangians are novel observations 
for the supersymmetric model of a free spinning relativistic particle.

Our present paper is organized as follows. In Sec. 2, we set up the notations and conventions for our
paper by discussing the bare essentials of (super)gauge transformations for the supersymmetric system
of massless spinning relativistic particle. Our Sec. 3 is devoted to the derivation of  (anti-)BRST
transformations of super-gauge (fermionic) variable $\chi (\tau)$ and its associated bosonic
(anti-)ghost variables $(\bar \beta) \beta$. An off-shoot of this exercise is the derivation of
(anti-)BRST transformations for the fermionic Lorentz vector variable $\psi_\mu (\tau)$.
Section 4 deals with the discussion of supersymmetric version of HC which enables us to derive
the (anti-)BRST symmetry transformations for the bosonic gauge variable $e(\tau)$ and its associated
fermionic (anti-)ghost variables $(\bar c)c$. A by-product of this exercise is the derivation of
 (anti-)BRST symmetry transformations for the bosonic target space variable $x_\mu (\tau)$.
We discuss the (anti-)BRST invariance of the CF-type  restriction, within the framework
of superfield formalism, in Sec. 5. Our Sec. 6 is devoted to the derivation of coupled
 Lagrangians and the proof of their (anti-)BRST invariance within 
the framework of superfield formalism. Finally, we make some concluding remarks in Sec. 7.

In our Appendices A and B, we discuss a set of interesting Lagrangians but show that they are {\it not}
appropriate in one way or the other ways. Our Appendix C is devoted to a brief discussion of a free massive
spinning relativistic particle within the framework of superfield approach to BRST formalism where we derive
some specific (anti-)BRST symmetries.

\section {Preliminaries: (super)gauge transformations}

Let us begin with the following first-order Lagrangian for the one (0 + 1)-dimensional 
(1D) supersymmetric model of a massless spinning relativistic particle [11]
\begin{eqnarray}
L_0 = p_\mu\; \dot x^\mu - \frac{e}{2}\; p^2 + \frac{i}{2}\; \psi_\mu \;\dot \psi^\mu  
+ i \;\chi\; (p_\mu\; \psi^\mu),
\end{eqnarray}
where the constraints $p^2 \approx 0$ and $p_\mu \psi^\mu \approx 0$ are first-class in 
nature (according to Dirac's prescription for the classification scheme [12,13]) and they have 
been incorporated in the above Lagrangian $L_0$ through the Lagrange multiplier variables 
$e(\tau)$ and $\chi(\tau)$ which are the analogs of the vierbein and Rarita-Schwinger 
(gravitino) fields of the 4D supergravity theories. In our present theory, these variables are
also the analogs of the gauge fields of the usual 4D supersymmetric gauge theories. 
Here the super world-line, traced out by the motion of the massless relativistic particle,
is parameterized by the monotonically increasing parameter $\tau$. This world-line is embedded
in a $D$-dimensional Minkowskian flat target supermanifold characterized by the target even (bosonic)
position variable $x_\mu (\tau) \;(\mu = 0, 1, 2,...,D-1)$ and odd (fermionic) spin variable 
$\psi_\mu (\tau)\; (\mu = 0, 1, 2,...,D-1)$ which are superpartners of each-other. The fermionic 
variables of the theory anticommute with one-other (i.e. $\psi_\mu^2 = 0,\; \chi^2 = 0, \;\psi_\mu \psi_\nu 
+ \psi_\nu \psi_\mu = 0, \; \psi_\mu \chi + \chi \psi_\mu = 0$, etc.). The conjugate momenta 
of the target space variables $x_\mu$ are $p^\mu = (\partial L_{0}/\partial \dot {x}_\mu)$ and 
$\dot {x}^\mu = (dx^\mu/d\tau) = e \;p^\mu - i\; \chi\; \psi^\mu,$ $\dot {\psi}^\mu = (d\psi^\mu/d\tau) 
= \chi \;p^\mu$ are the generalized ``velocities'' for the massless relativistic spinning particle 
corresponding to the above mentioned superpartners $x_\mu$ and $\psi_\mu$.

The above starting Lagrangian respects the local, continuous and infinitesimal (super)gauge 
transformations $(\delta_{(s)g})$ as follows (see, e.g. [11])
\begin{eqnarray}
&&\delta_{sg}\; x_\mu = \kappa\; \psi_\mu, \quad \delta_{sg}\; p_\mu = 0, \quad \delta_{sg} 
\;\psi_\mu = i\;\kappa\; p_\mu,\quad
\delta_{sg}\; \chi = i\; \dot \kappa, \quad \delta_{sg} \;e = 2\;\kappa\; \chi,\nonumber\\
&&\delta_g\; x_\mu = \xi\; p_\mu, \quad\quad \delta_g\; p_\mu = 0, \quad\quad \delta_g \;\psi_\mu = 0,\quad\quad
\delta_g\; \chi = 0, \quad\quad \delta_g \;e = \dot\xi,
\end{eqnarray}
where $(\kappa)\xi$ are the (fermionic)bosonic (super)gauge parameters. One can check explicitly that the Lagrangian $L_0$ 
transforms to a total derivative under $\delta_{(s)g}$. It will be noted that the (super)gauge transformations
are generated by the primary and secondary first-class constraints of the theory. Furthermore, the infinitesimal 
transformations $\delta_{sg}$ are nothing but the normal supersymmetric transformations for our present supersymmetric system.

The above transformations can be combined together (i.e. $\delta = \delta_g + \delta_{sg}$).
The resulting transformations $(\delta)$ for the dynamical variables of our present theory are
\begin{eqnarray}
\delta\; x_\mu = \xi\; p_\mu  + \kappa\; \psi_\mu, \quad \delta\; p_\mu = 0, \quad \delta \;\psi_\mu = i\;\kappa\; p_\mu,\quad
\delta\; \chi = i\; \dot \kappa, \quad \delta \;e = \dot \xi + 2\;\kappa\; \chi.
\end{eqnarray}
Under the above local, continuous and infinitesimal transformations, the Lagrangian $L_0$ transforms to a total derivative as follows
\begin{eqnarray}
\delta L_0 = \frac {d}{d\tau} \;\Big[ \frac {\xi}{2} \;p^2 + \frac {\kappa}{2} \;(p \cdot \psi)\Big],
\end{eqnarray} 
where, for the sake of brevity, we have chosen $p_\mu \psi^\mu = p \cdot \psi$. Henceforth, we shall follow
this notation in the whole body of our present text. It is clear that the action integral
$S = \int d\tau L_0$ remains invariant under the transformations (4) for the physically well-defined dynamical variables
of the theory which vanish off at infinity. The limiting cases of (3) and (4) produce results for the transformations
(2), separately and independently.

We close this section with the following remarks. First, the BRST transformations corresponding to the gauge 
transformations ($\delta_g$) have been written in [11] which are nilpotent of order two. Second, it has been 
pointed out in [11] that the BRST-type transformations exist corresponding to the supergauge transformations ($\delta_{sg}$)
but they are {\it not} nilpotent of order two. Third, the BT superfield formalism has been applied to
derive the proper nilpotent (anti-)BRST symmetry transformations corresponding to ($\delta_{(s)g}$)
separately and independently [14]. However, the (anti-)BRST transformations, corresponding to
$\delta$, have {\it not} yet been derived within the framework of superfield formalism.
Lastly, the BRST transformations, corresponding to the transformations 
$\delta = \delta_g + \delta_{sg}$, have been mentioned in [11,14] which are found to be nilpotent of order two. The {\it proper} (i.e. nilpotent
and absolutely anticommuting) anti-BRST transformations for the above transformations have, however, {\it not} been quoted in [11,14].
In our forthcoming sections, we attempt to resolve these contagious issues in a systematic manner by applying the key concepts
of superfield formalism proposed in [4,5] and modify it consistently for the derivation of the correct
nilpotent and absolutely anticommuting (anti-)BRST transformations (in the case of our supersymmetric system 
where the bosonic as well as fermionic variables co-exist together).

\section {Horizontality condition: (anti-)BRST transformations for variables $\chi(\tau),\;
\beta(\tau),\; \bar\beta(\tau), \; \psi_\mu(\tau)$}

It is elementary to note that the supergauge variable $\chi(\tau)$ is a fermionic auxiliary variable in the theory
(described by the Lagrangian $L_0$) as there is no kinetic term for 
this variable (because $\dot {\chi}^2 = 0$). The common folklore in the gauge theory states that
the kinetic term of a gauge variable is hidden in the curvature term which owes its origin
to the exterior derivative $d$ (with $d^2 = 0$). For our present 1D theory, the curvature 2-form,
in the language of the exterior derivative
$d = d\tau \;\partial_\tau$ and 1-form fermionic connection $f^{(1)} = d\tau\; \chi(\tau)$ for 
the variable $\chi(\tau)$, is $F^{(2)} = d\; f^{(1)} =(d\tau \wedge d\tau)\;\partial_\tau\; \chi (\tau) \equiv 
(d\tau \wedge d\tau)\; \dot {\chi} (\tau) = 0$ (due to the basic property of wedge product $d\tau \wedge d\tau = 0$).

To derive the (anti-)BRST transformations for the supergauge variable $\chi(\tau)$ and its associated (anti-)ghost
fields $(\bar\beta)\beta,$ we consider the supergauge theory on a (1, 2)-dimensional supermanifold where the
exterior derivative $d = d\tau\;\partial_\tau$ and 1-form fermionic connection $f^{(1)} = d\tau\;\chi(\tau)$
are generalized onto the (1, 2)-dimensional supermanifold as [4,5,14]
\begin{eqnarray}
&&d \rightarrow  \tilde d = dZ^M \partial_M = d\tau \;\partial_\tau + d\theta \;\partial_\theta  
+ d \bar\theta \;\partial_{\bar\theta}, \nonumber\\
&&f^{(1)}\rightarrow  \tilde {\cal F}^{(1)} =dZ^M \tilde A_M (\tau, \theta, \bar\theta) 
= d\tau \;K(\tau, \theta, \bar\theta) + i\;d\theta \;\bar{\cal B} (\tau, \theta, \bar\theta)  
+ i\;d \bar\theta \; {\cal B}(\tau, \theta, \bar\theta), 
\end{eqnarray}
where $Z^M = (\tau, \theta, \bar\theta)$ and $\partial_M = (\partial_\tau, \partial_\theta, \partial_{\bar\theta})$
are the superspace coordinates and superspace derivatives that characterize the (1, 2)-dimensional supermanifold and supermultiplet
vector superfield $\tilde A_M  (\tau, \theta, \bar\theta) \equiv  [K(\tau, \theta, \bar\theta), {\cal B}(\tau, \theta, \bar\theta), 
\bar{\cal B}(\tau, \theta, \bar\theta)]$.
Here $\tau$ is an even (bosonic) element and ($\theta, \bar\theta$) are the odd (fermionic)
elements of the Grassmann algebra and superfield $K $ is fermionic 
and (${\cal B}, \bar{\cal B}$) are bosonic in nature. The superfields $K (\tau, \theta, \bar\theta),\;{\cal B}(\tau, \theta, \bar\theta)$
and $\bar{\cal B}(\tau, \theta, \bar\theta)$ can be expanded along the Grassmannian directions ($\theta, \bar\theta$)
of the (1, 2)-dimensional supermanifold in the following manner (see, e.g. [4,5,14])
\begin{eqnarray}
&&K (\tau, \theta, \bar\theta) = \chi (\tau) + \theta\; \bar b_1 (\tau) + \bar\theta\; b_1 (\tau) 
+  \theta\;\bar\theta\; f_1(\tau),\nonumber\\ &&{\cal B}(\tau, \theta, \bar\theta) = \beta (\tau) 
+ \theta\; \bar f_2 (\tau) + \bar\theta\; f_2 (\tau) + i\; \theta\;\bar\theta\; b_2(\tau),\nonumber\\
&&\bar{\cal B}(\tau, \theta, \bar\theta) = \bar\beta (\tau) + \theta\; \bar f_3 (\tau) 
+ \bar\theta\; f_3 (\tau) + i\; \theta\;\bar\theta\; b_3(\tau).
\end{eqnarray}
It is evident that, in the limiting case $(\theta, \bar\theta)\rightarrow 0$, we retrieve the basic
variables $\chi(\tau), \beta(\tau)$ and $\bar\beta(\tau)$, respectively, of the 1D (anti-)BRST invariant 
theory 
of a free spinning particle. In the above, the sets ($ \chi, f_1, \bar f_2, f_2, \bar f_3, f_3$) 
and ($\bar b_1, b_1, \beta, \bar\beta, b_2, b_3$) are fermionic and bosonic in nature, respectively, 
and their equality (in numbers) ensures the existence of supersymmetry (SUSY) in the theory. The variables 
($b_1, \bar b_1, f_1, \bar f_2, f_2, b_2, \bar f_3, f_3, b_3$) are secondary
and they have to be determined in terms of the basic and auxiliary dynamical variables of the (anti-)BRST invariant
1D theory of spinning particle by the application of the usual HC.

We apply now the standard technique of HC which physically requires that the gauge (and/or BRST) invariant curvature 2-form
$F^{(2)} = d\;f^{(1)}$ has to be independent of the Grassmannian (odd) variables $\theta, \bar\theta$ of the
superspace coordinate $Z^M$ (and the corresponding differentials) in the following relationship
\begin{eqnarray}
\tilde d \;\tilde {\cal F}^{(1)} = d \; f^{(1)} = 0,  
\end{eqnarray}
where the explicit expression for the l.h.s., in terms of the multiplet superfields, is
\begin{eqnarray}
\tilde d \;\tilde {\cal F}^{(1)} &=& (d\tau \wedge d\theta) \;(i\; \partial_{\tau} \bar {\cal B} - \partial_{\theta} K) 
- i\; (d\theta \wedge d\theta)\;(\partial_{\theta} \bar{\cal B}) + (d\tau \wedge d\bar\theta)\;(i\; \partial_{\tau} {\cal B} 
- \partial_{\bar\theta} K)\nonumber\\ &-& i\; (d\theta \wedge d\bar\theta) \;(\partial_\theta {\cal B} + \partial_{\bar\theta} 
\bar{\cal B}) - i\; (d\bar\theta \wedge d\bar\theta) \;(\partial_{\bar\theta} {\cal B}).
\end{eqnarray}
The above equality of HC, in (7), yields:
\begin{eqnarray}
\partial_{\theta} \bar{\cal B} = 0, \qquad \partial_{\bar\theta} {\cal B} = 0, \;\quad
\partial_{\bar\theta} \bar{\cal B} + \partial_{\theta} {\cal B} = 0, \;\quad 
\partial_{\theta} K = i\; \partial_{\tau} \bar{\cal B},\;\quad
\partial_{\bar\theta} K = i\; \partial_{\tau} {\cal B}.
\end{eqnarray}
The substitution of the expansions (6), leads to the following relations
\begin{eqnarray}
b_2 = b_3 = 0,\quad\qquad f_2 = \bar f_3 = 0, \quad\qquad f_3 + \bar f_2 = 0,
\end{eqnarray}
from the first {\it three} conditions of (9). If we choose $f_3 = \gamma$, it implies that $\bar f_2 = -\gamma$.
Thus, after the application of HC, we obtain the following expansions [4,5,14]
\begin{eqnarray}
{\cal B}^{(h)}(\tau, \theta, \bar\theta) & = &\beta (\tau) + \theta\; \Big(-i\;\gamma (\tau) \Big) + \bar\theta\; \Big(\;0 \;\Big) 
+ \theta\;\bar\theta\; \Big( \;0 \;\Big)\nonumber\\
 &\equiv &   \beta(\tau) + \theta \;\Big (s_{ab}\; \beta(\tau)\Big ) + \bar \theta \;\Big (s_b\; \beta(\tau)\Big ) + \theta\;\bar\theta \;\Big (s_b\;s_{ab}\; \beta(\tau)\Big ),\nonumber\\
\bar{\cal B}^{(h)}(\tau, \theta, \bar\theta) &=& \bar\beta (\tau ) + \theta\; \Big(\;0\; \Big)+ \bar\theta\; \Big(i\;\gamma (\tau ) \Big) +  \theta\;\bar\theta\; \Big(\;0\; \Big)\nonumber\\
 &\equiv & \bar\beta(\tau) + \theta\; \Big( s_{ab} \;\bar \beta (\tau) \Big)
+ \bar\theta\; \Big( s_b \;\bar \beta (\tau) \Big) + \theta\;\bar\theta\; \Big( s_b\; s_{ab} \;\bar \beta (\tau) \Big),
\end{eqnarray}
where the superscript ($h$) on the superfields denotes the expansion of these superfields after the application of HC. 
The above expansions would turn out to be very useful later on.

Now, the stage is set to derive the nilpotent (anti-)BRST symmetry transformations for the supergauge variable $\chi (\tau)$. 
In this connection, we can exploit the expansions (11) in the relationship (9) to obtain the secondary variables
of $K(\tau, \theta, \bar\theta)$ as 
\begin{eqnarray}
b_1 (\tau) = i\; \dot \beta (\tau), \quad\qquad {\bar b}_1 (\tau) = i\; \dot {\bar\beta}(\tau), \quad\qquad f_1(\tau) = - \dot \gamma (\tau).
\end{eqnarray}
Thus, the expansion of $K(\tau, \theta, \bar\theta)$, after the application of HC, is
\begin{eqnarray}
K^{(h)} (\tau, \theta, \bar\theta) &=& \chi (\tau) + \theta\; \Big(i\;\dot {\bar\beta} (\tau)\Big) +  \bar\theta\; \Big(i\; \dot{\beta} (\tau)\Big)
 +  \theta\;\bar\theta\; \Big (- \dot {\gamma} (\tau ) \Big)\nonumber\\
 &\equiv & \chi (\tau) + \theta\;\Big( s_{ab}\; \chi (\tau) \Big) + \bar\theta\;\Big( s_b \;\chi(\tau)\Big)
 +  \theta\;\bar\theta \;\Big(s_b\;s_{ab} \;\chi(\tau) \Big).
\end{eqnarray}
The relation $\dot {\psi}_\mu = \chi\; p_\mu$ (that emerges as the equation of motion from $L_0$) is a super-gauge
invariant quantity [cf. (2)] on the on-shell where $\dot {p}_\mu = 0$. Within the framework of augmented
BT superfield formalism [8-10,14], such relations should remain invariant on the supersymmetric (1, 2)-dimensional supermanifold. Thus, we have
the following supergauge invariant restriction (SGIR) on the superfields of the theory: 
\begin{eqnarray}
\dot {\Psi}_\mu (\tau, \theta, \bar\theta) - K^{(h)} (\tau, \theta, \bar\theta) \;P_\mu(\tau, \theta, \bar\theta)
= \dot {\psi}_\mu (\tau) - \chi(\tau)\; p_\mu (\tau ) = 0.
\end{eqnarray}
However, as we have seen, the momentum operator $p_\mu$ is a gauge invariant quantity
[cf. (2)]. Thus, we have the gauge invariant restriction (GIR) $P_\mu(\tau, \theta, \bar\theta) = p_\mu (\tau)$ where
\begin{eqnarray}
P_\mu (\tau, \theta, \bar\theta) = p_\mu (\tau) + \theta\; \bar F^{(1)}_\mu (\tau) + \bar\theta\; F^{(1)}_\mu (\tau)
 +  \theta\;\bar\theta \;B^{(1)}_\mu(\tau).
 \end{eqnarray}
 The equality $P_\mu(\tau, \theta, \bar\theta) = p_\mu (\tau)$, however, implies that we have $\bar F^{(1)}_\mu = F^{(1)}_\mu 
= B^{(1)}_\mu = 0$. As a consequence, the explicit expansion for the gauge invariant quantity 
$P_\mu (\tau, \theta, \bar\theta)$ is [4,5]
\begin{eqnarray}
P^{(g)}_\mu (\tau, \theta, \bar\theta) &=& p_\mu (\tau) + \theta\; \Big (\; 0\; \Big) + \bar\theta\; \Big 
(\; 0 \; \Big) +  \theta\;\bar\theta \;\Big (\; 0 \; \Big) \nonumber\\
 &\equiv& p_\mu (\tau) + \theta\; (s_{ab}\; p_\mu)  + \bar\theta\; (s_b\; p_\mu) 
+  \theta\;\bar\theta \; (s_b \;s_{ab}\; p_\mu),
\end{eqnarray}
which shows that $s_{(a)b}\; p_\mu = 0$. It is evident that the target space momenta $p_\mu$ are trivially (anti-)BRST
invariant quantities (as there are no transformations for them).
Thus, the above equation (14) can be re-expressed as:
\begin{eqnarray}
\dot {\Psi}_\mu (\tau, \theta, \bar\theta) = K^{(h)} (\tau, \theta, \bar\theta) \;p_\mu(\tau).
\end{eqnarray}
Exploiting the expansion (13) for $ K^{(h)} (\tau, \theta, \bar\theta)$ and taking the super-expansion 
of $\Psi_\mu (\tau, \theta, \bar\theta)$, along the Grassmannian directions of the (1, 2)-dimensional supermanifold, as
\begin{eqnarray}
\Psi_\mu (\tau, \theta, \bar\theta) = \psi_\mu (\tau) + \theta\; \bar b_\mu (\tau) + \bar\theta\; b_\mu (\tau) +  \theta\;\bar\theta\; f_\mu(\tau),
\end{eqnarray}
we obtain the secondary variables of $\Psi_\mu(\tau, \theta, \bar\theta)$ as: 
$b_\mu = i\; \beta\; p_\mu, \;\bar b_\mu = i\; \bar \beta\; p_\mu, \;f_\mu = - \gamma\; p_\mu.$
As a consequence, we have the following expansion for  $\Psi_\mu (\tau, \theta, \bar\theta)$, after the application of SGIR (14), 
along the Grassmannian directions of the (1, 2)-dimensional supermanifold
\begin{eqnarray}
\Psi_\mu^{(sg)} (\tau, \theta, \bar\theta) &=& \psi_\mu (\tau) + \theta\; \Big (i\;\bar\beta \; p_\mu \Big ) 
+ \bar\theta\; \Big( i\; \beta\; p_\mu \Big) +  \theta\;\bar\theta \;\Big (- \gamma \; p_\mu \Big) \nonumber\\
&\equiv & \psi_\mu (\tau) + \theta\; \Big ( s_{ab}\; \psi_\mu \Big ) 
+ \bar\theta\; \Big( s_b \; \psi_\mu \Big) +  \theta\;\bar\theta \;\Big (s_b\; s_{ab} \; \psi_\mu \Big),
\end{eqnarray}
where the superscript ($sg$) denotes the expansion after the application of SGIR [cf. (14)].

Taking the help of expansions in (11), (13), (16) and (19), we obtain the following off-shell nilpotent
and absolutely anticommuting (anti-)BRST symmetry transformations:
\begin{eqnarray}
&&s_{ab}\; \beta = - i\; \gamma,  \;\;\qquad\quad s_{ab}\; \gamma = 0, \qquad \qquad\;
s_{ab}\; \bar\beta = 0, \quad\quad\quad  s_{ab}\; \chi = i \;\dot {\bar \beta},\nonumber\\
&& s_{ab} \;\psi_\mu = i\; \bar \beta p_\mu, \;\quad\quad s_{ab}\; p_\mu = 0, \qquad \qquad
s_b \;\chi = i\;\dot \beta, \qquad \qquad s_b\; \beta = 0,  \nonumber\\
&&  s_b \;\bar \beta = i\; \gamma, \;\qquad\quad  s_b \;\psi_\mu = i\;\beta \;p_\mu, \qquad \qquad
s_b\; \gamma = 0,\; \;\qquad \quad s_b\; p_\mu = 0.
\end{eqnarray}
We can verify easily that $s^2_{(a)b} = 0$ and $s_b \;s_{ab} + s_{ab} \;s_b = 0$. 
In other words, it is clear that the two consecutive 
applications of $s_{(a)b}$ results in zero and the operator form of anticommutator 
(i.e. $\{s_b, s_{ab}\}$), acting on any variable, produces zero result.

We close this section with the remarks that the idea of HC leads to the derivation 
of  off-shell nilpotent (anti-)BRST symmetry transformations {\it only} for the gauge and 
its associated (anti-)ghost variables of the theory. However, these inputs help in 
the determination of (anti-)BRST symmetries for the other variables of the theory when we demand the
additional restriction, within the framework of augmented superfield formalism [8-10,14], where the
(super)gauge invariant quantities are also required to be independent of these Grassmannian variables.
For instance, we have been able to obtain the (anti-)BRST symmetry transformations
for the $\psi_\mu(\tau)$ variable by demanding the additional restriction [cf. (14)] on the 
(1, 2)-dimensional supermanifold because $\dot{\psi}_\mu = \chi\; p_\mu$ is a supergauge invariant quantity.
Thus, the importance of (S)GIR in our work is very decisive and crucial.

\section{Supersymmetrization of HC: (anti-)BRST symmetry transformations for $e(\tau),\; c(\tau),\; \bar c(\tau), \; x_\mu(\tau)$}

The HC plays a central role in the BT superfield approach to BRST formalism when it is applied to the BRST 
analysis of gauge theories [4,5]. However, our present model is a supersymmetric model where, for the 
first-time, the ideas of BT superfield formalism is being applied. The bosonic gauge dynamical variable 
of our present theory is $e (\tau)$. Thus, the 1-form ($A^{(1)}$), is defined in terms of it as
 $A^{(1)} = d\tau \;e(\tau)$. The curvature tensor turns out to be zero for this gauge dynamical variable because
$d \;A^{(1)} = 0$ (due to $d\tau \wedge d\tau = 0$). This is, however, only the bosonic part of the supersymmetric theory.
The total super 2-form curvature (which includes both the bosonic and fermionic parts) is the one where $e(\tau)$ and $\chi (\tau)$
should be present in a consistent and clear fashion, namely;
\begin{eqnarray}
 d \;A^{(1)} + i \;(f^{(1)} \wedge f^{(1)}) = 0,
\end{eqnarray}
where the fermionic 1-form $f^{(1)} = d\tau\; \chi(\tau)$ has already been discussed earlier\footnote{We note
that the above equation looks exactly like the well-known Maurer-Cartan equation which defines the curvature
tensor for the $SU(N)$ non-Abelian gauge theories. However, in such usual non-Abelian theories, there are no fermionic 
1-form connections as we have in our present analysis.}.

We have to generalize the relation (21) onto the (1, 2)-dimensional supermanifold and demand its equality to
itself as given below
\begin{eqnarray}
&&{\tilde d} \;{\tilde A^{(1)}} + i \;(\tilde{\cal F}^{(1)}_{(h)}\wedge\tilde{\cal F}^{(1)}_{(h)}) 
= d \;A^{(1)} + i \;( f^{(1)}\wedge f^{(1)}) = 0.
\end{eqnarray}
The above restriction is the generalization of the ordinary HC to a supersymmetric HC (i.e. SUSY-HC). The individual symbols on the l.h.s.
of (22) are
\begin{eqnarray}
&& \tilde d = d\tau \;\partial_\tau + d\theta \;\partial_\theta  + d \bar\theta \;\partial_{\bar\theta}, \nonumber\\
&& \tilde A^{(1)} = d\tau \;E(\tau, \theta, \bar\theta) + d\theta \;\bar F (\tau, \theta, \bar\theta) 
 + d \bar\theta \;F (\tau, \theta, \bar\theta), \nonumber\\
&& \tilde {\cal F}^{(1)}_{(h)} = d\tau \;K^{(h)}(\tau, \theta, \bar\theta) + i \;d\theta \;\bar {\cal B}^{(h)} (\tau, \theta, \bar\theta)  
+ i\; d \bar\theta\; {\cal B}^{(h)} (\tau, \theta, \bar\theta),
\end{eqnarray}
where  $K^{(h)}(\tau, \theta, \bar\theta), \; \bar {\cal B}^{(h)} (\tau, \theta, \bar\theta), 
\; {\cal B}^{(h)} (\tau, \theta, \bar\theta)$ are the superfields obtained after the application of HC (cf. Sec. 3).
It will be noted that the superfield $E(\tau, \theta, \bar\theta)$ is bosonic and the pair [$\bar F (\tau, \theta, \bar\theta), \;F (\tau, \theta, \bar\theta)$] is fermionic in $\tilde A^{(1)}$. 
As has been already mentioned, the r.h.s. of (22) is zero
on its own. The l.h.s. of (22) is explicitly expressed as follows:
\begin{eqnarray}
&&{\tilde d} \;{\tilde A}^{(1)} + i\; ({\tilde{\cal F}}^{(1)}_{(h)}\wedge{\tilde{\cal F}}^{(1)}_{(h)})
= (d\tau\wedge d\theta)\;(\partial_\tau \bar F - \partial_\theta E + 2\; \bar {\cal B}^{(h)}K^{(h)}) \nonumber\\
&&+(d\tau\wedge d\bar\theta)\;(\partial_\tau F - \partial_{\bar\theta} E + 2\; {\cal  B}^{(h)}K^{(h)}) 
- (d \theta \wedge d\bar\theta)\; (\partial_\theta F + \partial_{\bar\theta} \bar F + 2\;i\;{\cal B}^{(h)}\;
\bar {\cal B}^{(h)})
\nonumber\\ &&- (d\theta \wedge d \theta)\;(\partial_{\theta} \bar F 
+ i \; {\bar {\cal B}}^{(h)} \bar {\cal B}^{(h)}) -
(d \bar \theta \wedge d \bar \theta)\;(\partial_{\bar \theta}  F + i \;  {\cal B}^{(h)}  {\cal B}^{(h)}).
\end{eqnarray}
The restriction in (22) implies that the coefficients of $(d\tau \wedge d\theta), (d\tau \wedge d\bar\theta), 
(d\theta \wedge d\bar\theta), (d\theta \wedge d\theta), (d\bar\theta \wedge d\bar\theta)$
have to be set equal to zero. These requirements lead to the following  
\begin{eqnarray}
&&\partial_\tau \bar F - \partial_\theta E + 2 \;\bar {\cal B}^{(h)}K^{(h)} = 0,\quad \partial_\tau F
- \partial_{\bar\theta} E + 2\; {\cal  B}^{(h)}K^{(h)} = 0,\nonumber\\ 
&&\partial_{\theta} F + \partial_{\bar\theta} \bar F + 2\; i \; {\bar {\cal B}}^{(h)} {\cal B}^{(h)} = 0,
\quad\qquad \partial_{\bar \theta}  F + i \;  {\cal B}^{(h)}  {\cal B}^{(h)}= 0, \nonumber\\ && \partial_{\theta}  \bar F 
+ i \; {\bar {\cal B}}^{(h)} \bar {\cal B}^{(h)} = 0.
\end{eqnarray}
The above relationships allow the derivation of secondary variables in terms of the auxiliary 
and dynamical variables of the BRST-invariant 1D theory of spinning relativistic particle when 
the proper expansions of superfields are plugged in the above equation (25).

To achieve the above theoretically important goals, first of all, we have to expand the superfields 
$E(\tau,\theta, \bar\theta),\; F(\tau, \theta, \bar\theta)$ and $\bar F(\tau, \theta, \bar\theta)$ 
along the Grassmannian directions as: 
\begin{eqnarray}
&& E(\tau, \theta, \bar \theta) = e(\tau) + \theta\; \bar f(\tau) + \bar \theta\; f(\tau) 
+ i\;\theta\;\bar\theta\; B(\tau), \nonumber\\ && F(\tau, \theta, \bar \theta) = c(\tau) 
+ i \;\theta\; \bar b_1(\tau) + i\; \bar \theta\; b_1(\tau) + i\;\theta\;\bar\theta\; s_1(\tau),
\nonumber\\ && \bar F(\tau, \theta, \bar \theta) = \bar c(\tau) + i \;\theta\; \bar b_2(\tau) 
+ i\; \bar \theta\; b_2(\tau) + i\;\theta\;\bar\theta\; \bar s_1(\tau),
\end{eqnarray} 
where the basic tenet of SUSY is satisfied accurately because the number of bosonic variables 
($e, B, \bar b_1, b_1, \bar b_2, b_2$) and fermionic variables ($\bar f, f, c, \bar c, s_1, \bar s_1$) 
match. We have to  determine all the secondary variables ($f, \bar f, B, b_1, \bar b_1, b_2, \bar b_2, s_1, \bar s_1$)
in terms of the auxiliary and basic variables of the BRST invariant 1D model of massless spinning relativistic particle.

The last three relationships of (25) yield the following:
\begin{eqnarray} 
b_1 = - \beta^2,\quad s_1 = -2\; i\; \beta\;\gamma, \quad \bar b_2 = -\bar \beta^2, 
\quad \bar s_1 = - 2\;i\; \bar \beta\; \gamma, \quad \bar b_1 + b_2 = - 2\;\bar \beta\; \beta.
\end{eqnarray}
As a consequence, we have the following expansions for the fermionic superfields [4,5]
\begin{eqnarray} 
F^{(sh)}(\tau, \theta, \bar\theta) &=& c +  \theta\; \Big(i \;\bar b_1\Big)+  \bar\theta \;\Big(-i\; {\beta}^2\Big)
 + \theta\;\bar\theta\; \Big( 2\;\beta\;\gamma \Big) \nonumber\\
&\equiv & c +  \theta \;\Big(s_{ab} \;c \Big) +  \bar\theta\; \Big(s_b\; c\Big) + \theta\;\bar\theta \;\Big(s_b\; s_{ab}\; c \Big), \nonumber\\ 
\bar F^{(sh)}(\tau, \theta, \bar \theta) &=& \bar c +  \theta\;\Big(-i \;{\bar \beta}^2 \Big) +  \bar\theta\; \Big(i\; b_2\Big) 
+ \theta\;\bar\theta\; \Big(2 \;\bar\beta\;\gamma \Big)\nonumber\\
&\equiv & \bar c +  \theta \;\Big(s_{ab}\;\bar c \Big) +  \bar\theta\; \Big(s_b\; \bar c\Big) + \theta\;\bar\theta\;\Big(s_b \;s_{ab}\;\bar c\Big),
\end{eqnarray}
where the superscript ($sh$) on the superfields denotes the expansions of the superfields after the application of SUSY-HC. 
Identifying $\bar b_1 = \bar b,\; b_2 = b$, we obtain the following (anti-)BRST transformations
\begin{eqnarray} 
s_b \;c = - i\; \beta^2, \qquad s_b\; {\bar c} = i\; b, \qquad 
s_{ab}\; \bar c = - i\; {\bar \beta}^2, \qquad s_{ab}\; c = i\; \bar b, 
\end{eqnarray}
which are off-shell nilpotent of order two (i.e. $s^2_{(a)b} = 0$) and absolutely anticommuting $s_b \;s_{ab} 
+ s_{ab} \;s_b = 0$
in nature because the following is sacrosanct, namely;
\begin{eqnarray} 
s_b\; \beta = 0, \qquad s_b\; b= 0, \qquad s_{ab}\; \bar\beta = 0, \qquad s_{ab}\; \bar b = 0, \quad b+ \bar b = - 2\; \beta\; \bar\beta.
\end{eqnarray}
The last entry in the above is nothing but the CF-type restriction which is very crucial for 
our further discussions and it emerges from, setting equal to zero, 
the coefficient of ($d\theta\wedge d\bar\theta$) in the restriction (22).
The first two relations of (25), with inputs from (28), (29) and 
(${\cal B}^{(h)},\;{\bar {\cal B}}^{(h)}, \; K^{(h)}$),
lead to the determination of secondary variables of $E(\tau,\theta, \bar\theta)$ as  
\begin{eqnarray}
f = \dot c + 2 \;\beta \;\chi, \quad \bar f = \dot{\bar  c} + 2\; \bar \beta\; \chi,  \quad
B = (\dot b + 2\;\bar \beta\; \dot\beta - 2\;\chi\; \gamma) \equiv - (\dot{\bar b} + 2\; \dot{\bar \beta}\; \beta + 2 \;\chi\; \gamma).
\end{eqnarray} 
As a consequence, we have the following expansion for this bosonic gauge superfield [4,5]
\begin{eqnarray}
E^{(sh)}(\tau, \theta, \bar \theta) &=& e + \theta\; \Big(\dot {\bar c} + 2\;\bar \beta\;\chi \Big) + \bar \theta\; \Big(\dot c + 2\;\beta\;\chi \Big) 
+ \theta\;\bar\theta\; \Big(i\; \dot b + 2\;i\; \bar \beta\;\dot \beta - 2\;i\; \chi\;\gamma \Big) \nonumber\\
&\equiv&  e + \theta \;\Big(s_{ab}\;e\Big) + \bar \theta\; \Big(s_{b}\;e \Big) + \theta\;\bar\theta\; \Big(s_b \;s_{ab}\;e\Big),
\end{eqnarray} 
which shows that $s_b \;e = \dot {c} + 2\;\beta\; \chi, \; s_{ab}\; e = \dot{\bar c} +2\;\bar\beta\; \chi, 
\; s_b \;s_{ab}\;e = i\; (\dot b + 2\;\bar \beta\;\dot \beta + 2\;\gamma\;\chi)$.

Now, we are in a position to derive the nilpotent (anti-)BRST symmetry transformations for the target space position variable
$x_\mu (\tau)$. It can be seen that $\dot {x}_\mu = e\; p_\mu - i\; \chi \;\psi_\mu$ is a supergauge invariant quantity
on the on-shell where $\dot {p}_\mu = 0$. As a consequence, we demand the invariance of this relationship on the
(1, 2)-dimensional supermanifold, namely;
\begin{eqnarray}
{\dot X}_\mu (\tau, \theta, \bar \theta) = E^{(sh)}(\tau, \theta, \bar \theta) \;p_\mu(\tau)
 - i \;K^{(h)}(\tau, \theta, \bar\theta) \;\Psi^{(sg)}_\mu (\tau, \theta, \bar \theta), 
\end{eqnarray} 
where we have taken $P^{(g)}_\mu(\tau, \theta, \bar\theta) = p_\mu(\tau)$ for the obvious reasons.
Taking the following general super-expansion for the target space superfield $X_\mu (\tau, \theta, \bar\theta)$ along the
Grassmannian directions of (1, 2)-dimensional supermanifold, namely;
\begin{eqnarray}
X_\mu (\tau, \theta, \bar \theta) = x_\mu (\tau) + \theta\; \bar R_\mu(\tau) 
+ \bar\theta  \;R_\mu (\tau)+ i\;\theta \;\bar\theta\; S_\mu (\tau),
\end{eqnarray} 
and exploiting the expressions for $E^{(sh)}(\tau, \theta, \bar \theta), \; K^{(h)}(\tau, \theta, \bar\theta)$ and 
$\Psi^{(sg)}_\mu (\tau, \theta, \bar \theta)$ from (32), (13) and (19), respectively, we obtain the following 
\begin{eqnarray}
&&{\dot {\bar R}}_\mu = {\dot {\bar c}}\; p_\mu + \dot {\bar \beta}\;\psi_\mu + \bar\beta\; \chi\; p_\mu, \qquad 
{\dot R}_\mu = {\dot c}\; p_\mu + \dot {\beta}\; \psi_\mu + \beta \;\chi\;  p_\mu, \nonumber\\
&&\dot { S}_\mu = (\dot {b_2}  + \dot {\bar \beta}\; \beta + \dot {\beta}\; \bar \beta- \chi\; \gamma)\; p_\mu + \dot{\gamma} \;\psi_\mu,          
\end{eqnarray} 
which have come out from, setting equal to zero, the coefficients of $\theta, \bar\theta$ and $\theta\bar\theta$.
Furthermore, we note that the l.h.s. of (35) is a total time derivative. Hence, the r.h.s. has to be also made 
a total time derivative so that both sides could be compared\footnote{It should be noted that
the comparison in (35), after the use of equations of motion $\dot p_\mu = 0, \dot \psi_\mu = \chi \; p_\mu$,
might differ by a constant factor w.r.t. the evolution parameter $\tau$. However, for the sake of simpliciy,
we have taken that constant equal to zero because physics would not be affected by this choice.}. 
In this context, it is interesting to
point out that the equations of motion $\dot p_\mu = 0$ and $\dot {\psi_\mu} = \chi\; p_\mu$ turn out to be
quite handy and, ultimately, we obtain: $
\bar R_\mu =  \bar c\; p_\mu + \bar \beta\; \psi_\mu,\;
R_\mu = c\; p_\mu + \beta \;\psi_\mu, \;
S_\mu = (b  + \bar \beta \;\beta)\; p_\mu + \gamma \;\psi_\mu$.
Substitution of these secondary variables, in the expansion of $X_\mu (\tau, \theta, \bar\theta)$,
leads us to obtain the following explicit expansion for the target space superfield 
\begin{eqnarray}
X^{(sg)}_\mu (\tau, \theta, \bar \theta) &=& x_\mu (\tau) + \theta \;\Big(\bar c\; p_\mu + \bar \beta\; \psi_\mu \Big) 
+ \bar\theta\; \Big(c\;p_\mu + \beta\; \psi_\mu \Big) \nonumber\\
&+& \theta\;\bar\theta \;\Big((i \; b  + i \;\bar \beta \;\beta)\; p_\mu + i\;\gamma \;\psi_\mu \Big) 
\nonumber\\ &\equiv& x_\mu(\tau) + \theta\; \Big(s_{ab} \;x_\mu \Big) + \bar\theta\; \Big(s_b\; x_\mu \Big)
+ \theta\;\bar\theta \Big(s_b\;s_{ab}\; x_\mu \Big),
\end{eqnarray} 
where the superscript ($sg$) stands for the superfield $X_\mu (\tau, \theta, \bar\theta)$
obtained after the implementation of the SGIR (33). As a result of (36), we obtain the following
off-shell nilpotent ($s^2_{(a)b} = 0$) (anti-)BRST symmetry transformations $s_{(a)b}$ for $x_\mu (\tau)$:
\begin{eqnarray} 
s_b \;x_\mu = c\;p_\mu + \beta\; \psi_\mu, \;\quad 
s_{ab} \;x_\mu = {\bar c} \;p_\mu + \bar \beta \;\psi_\mu, \;\quad s_b\;s_{ab} \;x_\mu = (i \; b  
+ i \;\bar \beta \;\beta)\; p_\mu + i\;\gamma \;\psi_\mu.
\end{eqnarray}
The nilpotency can be checked explicitly by taking the (anti-)BRST symmetry transformations for
 $c,\; p_\mu,\; \beta,\; \psi_\mu,\; \bar c,\; \bar\beta$ from our earlier equations [or, cf. (39), (40) below].

We wrap up this section with the statement that it is the requirement of nilpotency property
that we obtain the (anti-)BRST transformations like $s_b \; b = 0, \; s_b\; \beta = 0,
\; s_{ab}\; \bar b = 0, \; s_{ab}\; \bar\beta = 0$, etc. The sanctity of the absolute anticommutativity property 
also helps us in the determination of the (anti-)BRST symmetries for the Nakanishi-Lautrup type auxiliary 
variables like $s_b \; \bar b = -2\;i\;\beta\; \gamma$ and $s_{ab} \; b = +2\;i\;\bar\beta\; \gamma$, etc.

\section{CF-type restriction: (anti-)BRST invariance}

We have been able to obtain all the expansions of the superfields in terms of the basic and 
auxiliary variables of the (anti-)BRST invariant Lagrangian for our present supersymmetric model in
one ($0 + 1$)-dimension (1D) of spacetime. These expansions, after the application of HC, GIR, SGIR and SUSY-HC, 
can be written together in an explicit form as: 
\begin{eqnarray}
X^{(sg)}_\mu (\tau, \theta, \bar \theta) &=& x_\mu (\tau) + \theta\; \Big(\bar c\; p_\mu + \bar \beta\; \psi_\mu \Big) 
+ \bar\theta\; \Big(c\; p_\mu + \beta\; \psi_\mu \Big) \nonumber\\
&+& \theta\;\bar\theta \;\Big ((i \; b  + i \;\bar \beta\; \beta)\; p_\mu + i\;\gamma \;\psi_\mu \Big), \nonumber\\
E^{(sh)}(\tau, \theta, \bar \theta) &=& e(\tau) + \theta \;\Big(\dot {\bar c} + 2\; \bar \beta\;\chi \Big) 
+ \bar \theta\; \Big(\dot c + 2 \; \beta\;\chi \Big) \nonumber\\
&+& \theta\;\bar\theta\; \Big(i\; \dot b + 2\;i \;\bar \beta\; \dot{\beta} - 2\;i\; \chi\; \gamma\Big),\nonumber\\
\Psi^{(sg)}_\mu (\tau, \theta, \bar\theta )&=& \psi_\mu (\tau)+ \theta \;\Big(i\;\bar\beta\;p_\mu\Big)  + \bar \theta\;  \Big(i\;\beta\;p_\mu\Big)
 + \theta \;\bar \theta\;  \Big(\;- \gamma\;p_\mu\Big), \nonumber\\
K^{(h)}(\tau, \theta, \bar \theta) &=& \chi(\tau) + \theta \;\Big(i\; \dot {\bar \beta}\Big) 
+ \bar\theta \;\Big(i \;\dot\beta\Big) + \theta\;\bar\theta \;\Big(- \dot\gamma\Big), \nonumber\\
F^{(sh)}(\tau, \theta, \bar \theta) &=& c(\tau) +  \theta\; \Big(i\; \bar b\Big) +  \bar \theta\;\Big (-i \;{\beta}^2\Big)
 + \theta\;\bar\theta\;\Big(2\;\beta\;\gamma\Big), \nonumber\\
\bar F^{(sh)}(\tau, \theta, \bar \theta) &=& \bar c(\tau) +  \theta\;\Big(-i\; {\bar \beta}^2\Big) +  \bar \theta\; \Big(i\; b\Big) 
+ \theta\;\bar\theta\;\Big (2 \;\bar\beta\;\gamma\Big), \nonumber\\
{\cal B}^{(h)}(\tau, \theta, \bar \theta) &=& \beta(\tau) +  \theta\; \Big(- i\; \gamma\Big) + \bar\theta\; \Big (\; 0\;\Big) 
+\theta\;\bar\theta \; \Big(\; 0 \; \Big), \nonumber\\
\bar {\cal B}^{(h)}(\tau, \theta, \bar \theta) &=& \bar \beta(\tau) +  \theta\; \Big(\; 0 \; \Big) + \bar\theta\; \Big(i\; \gamma\Big) +
\theta\;\bar\theta\;\Big (\; 0 \; \Big), \nonumber\\
P^{(g)}_\mu (\tau, \theta, \bar\theta )&=& p_\mu (\tau)+ \theta \;\Big(\;0\;\Big)  + \bar \theta\;  \Big(\;0\;\Big)
 + \theta \;\bar \theta\;  \Big(\;0\;\Big).
\end{eqnarray}
A close look at the above expansions leads to the derivation of (anti-)BRST symmetry transformations $s_{(a)b}$,
in their full blaze of glory, as listed below:
\begin{eqnarray} 
&& s_{ab}\; x_\mu = {\bar c}\; p_\mu + \bar \beta \;\psi_\mu, \quad\qquad s_{ab}\; e = \dot {\bar c} + 2 \;\bar \beta\; \chi,  
\;\quad\qquad s_{ab} \;\psi_\mu = i \;\bar \beta\; p_\mu,\nonumber\\
&& s_{ab}\; \bar c = - i \;{\bar \beta}^2, \quad s_{ab}\; c = i\; \bar b, \quad s_{ab}\; \bar \beta = 0, 
\;\quad s_{ab} \; \beta = - i\; \gamma, \quad s_{ab}\; p_\mu = 0, \nonumber\\
&& s_{ab} \;\gamma = 0, \qquad\qquad s_{ab}\; \bar b = 0, \quad\qquad s_{ab}\;\chi = i\; \dot {\bar \beta}, 
\qquad\quad s_{ab} \; b =  2\; i\; \bar \beta\; \gamma,
\end{eqnarray}
\begin{eqnarray}
&&s_b\; x_\mu = c\;p_\mu + \beta \;\psi_\mu, \quad\qquad s_b\; e = \dot c + 2\;\beta\; \chi,  
\quad\qquad s_b\; \psi_\mu = i\;\beta\; p_\mu,\nonumber\\
&& s_b\;c = - i\; \beta^2, \;\quad s_b \;{\bar c} = i\; b, \;\quad s_b \;\beta = 0, 
\;\quad s_b \;\bar \beta = i \;\gamma, \;\quad s_b\; p_\mu = 0,\nonumber\\
&& s_b \;\gamma = 0, \qquad\quad s_b \;b = 0, \qquad\quad s_b \;\chi = i\; \dot \beta, 
\qquad\quad s_b\; \bar b = - 2\; i\; \beta\; \gamma.
\end{eqnarray}
It is elementary to check that the transformations in (39) and (40) are nilpotent of order two
(i.e. $s^2_{(a)b} = 0$) which establishes the fermionic nature of (anti-)BRST transformations.

Now we discuss a bit about the CF-type condition ($b + \bar b = - 2\; \beta \; \bar\beta$) of our present theory,
its importance and its (anti-)BRST invariance within the framework of superfield  formalism.
We re-emphasize that, from the SUSY-HC (22), we obtain the CF-type restriction ($b + \bar b + 2\; \beta \;
\bar\beta = 0$)
when we set equal to zero the coefficient of ($d\theta \wedge d\bar\theta$).
It can be checked explicitly that the operator form of $\{s_b, s_{ab}\} = s_b \;s_{ab} + s_{ab} \;s_b$, acting on all the variables 
of the present theory, is trivially zero except for  the following:
\begin{eqnarray}    
\{s_b, s_{ab}\}\;e (\tau) \neq 0, \;\;\qquad\qquad \;\;\{s_b, s_{ab}\}\;x_\mu (\tau) \neq 0.
\end{eqnarray}
Thus, it appears that one of the sacrosanct properties  of BRST formalism is lost. However, at this stage, the CF-type restriction
comes to our rescue. One can verify that there is an absolute anticommutativity in the 
 theory because the r.h.s. of (41) is also zero on a constrained super world-line, defined by the CF-type equation 
($b + \bar b + 2\; \beta \;\bar\beta = 0$).
We conclude that the absolute anticommutativity of the off-shell nilpotent (anti-)BRST symmetries is respected
in the theory because of the presence of CF-type restriction  which emerges from the augmented 
version of BT superfield formalism [8-10,14]. One of the key features of CF-type restriction is that 
it is an (anti-)BRST invariant quantity because
\begin{eqnarray}
s_{(a)b}\;[b + \bar b + 2\; \beta\; \bar\beta] = 0,
\end{eqnarray}
where we have used explicitly the transformations listed in (39) and (40). The above observation establishes the
fact that CF-type restriction is a physical restriction (in some sense) because it is an (anti-)BRST invariant quantity.

The (anti-)BRST invariance of CF-type restriction [cf. (42)] can be captured within the framework
of augmented BT superfield formalism as well. Towards this goal in mind, it can be seen [from our expansions
(38)] that we have the following generic relationship between the Grassmannian derivatives 
($\partial_\theta, \partial_{\bar\theta}$) of the (1, 2)-dimensional supermanifold and the nilpotent
(anti-)BRST transformations $s_{(a)b}$ on the 1D dynamical variables, namely;
\begin{eqnarray}
&&\lim_{\theta \to 0}  \;\frac {\partial}{\partial{\bar\theta}}
 \;\Omega^{(h, sh, g, sg)} (\tau, \theta, \bar\theta) = s_b \;w(\tau),\nonumber\\
&&\lim_{\bar\theta \to 0}  \;\frac {\partial}{\partial{\theta}}
 \;\Omega^{(h, sh, g, sg)} (\tau, \theta, \bar\theta) = s_{ab} \;w(\tau),\nonumber\\
&&\frac {\partial} {\partial {\bar\theta}}\;\frac {\partial}{\partial{\theta}}
 \;\Omega^{(h, sh, g, sg)} (\tau, \theta, \bar\theta) = s_b\;s_{ab} \;w(\tau),
\end{eqnarray}
where $\Omega^{(h, sh, g, sg)} (\tau, \theta, \bar\theta)$ are the generic expansions of superfields [cf. (38)],
derived after the application of (SUSY-)HC as well as (S)GIR and  $\omega(\tau)$ is the generic dynamical variable 
of the (anti-)BRST invariant 1D theory [cf. (51), (52) below] for the description of 
a free spinning relativistic particle [e.g. $\omega(\tau) = e(\tau),\; \chi(\tau), x_\mu(\tau),
\; \psi_\mu (\tau), \; c(\tau), \;\bar c (\tau)$, etc.].

In view of the mapping in (43), it can be seen that the coefficient of ($d\theta \wedge d\bar\theta$), which yields
CF condition [cf. (25), (30)], is independent of $\theta$ and $\bar\theta$. To see it clearly, first of all, 
we note the following [which have been already quoted in (25) and (30)], namely;
\begin{eqnarray}
\partial_{\theta} F^{(sh)} + \partial_{\bar\theta} \bar F^{(sh)}
+ 2 \;i \; {\bar {\cal B}}^{(h)} {\cal B}^{(h)} = 0 \; \quad \Longrightarrow \;\quad b + \bar b +2\;\beta \;\bar\beta = 0.
\end{eqnarray} 
Next step is to prove the BRST invariance of ($b +\bar b +2\;\beta \;\bar\beta = 0$), in the language of superfield formalism.
In this context, we observe very explicitly [in view of (43)] that
\begin{eqnarray}
\lim_{\theta \to 0} \; \partial_{\bar\theta}\;\Big[\partial_{\theta} F^{(sh)} + \partial_{\bar\theta} \bar F^{(sh)}
+ 2\; i \; {\bar {\cal B}}^{(h)} {\cal B}^{(h)} \Big] = 2\;\beta\; \gamma - 2\; \beta\; \gamma = 0.
\end{eqnarray}
Similarly, for the anti-BRST invariance of ($b +\bar b +2\;\beta \; \bar\beta = 0$), we can clearly check that
\begin{eqnarray}
\lim_{\bar\theta \rightarrow 0}\; \partial_{\theta}\;\Big[\partial_{\theta} F^{(sh)} + \partial_{\bar\theta} \bar F^{(sh)}
+ 2 \;i \; {\bar {\cal B}}^{(h)} {\cal B}^{(h)} \Big] = 2\;\bar\beta\; \gamma - 2\; \bar\beta\; \gamma = 0.
\end{eqnarray}
It is transparent that the equations (45) and (46) do capture (42). Thus, from the above equations,
it is clear that we have captured the (anti-)BRST invariance of CF-type restriction within the 
framework of our augmented version of BT superfield formalism.

One of the most important features of CF-type restriction is its role in the derivation of coupled 
(but equivalent) Lagrangians for a given theory. The beauty of these Lagrangians is the fact that they are 
(anti-)BRST invariant on the constraint surface (in our 1D case a super world-line) which is defined by 
the CF-type equation ($b + \bar b + 2\;\bar\beta\;\beta = 0$). Furthermore, such coupled Lagrangians yield
the CF-type restriction as an off-shoot of their Euler-Lagrange equations of motion. In our next section, 
we are going to discuss this aspect of the (anti-)BRST invariant CF-type restriction in the full blaze of its glory.

\section{Coupled Lagrangians: (anti-)BRST invariance}

We derive here the most appropriate coupled (but equivalent) Lagrangians for the spinning relativistic particle 
which respect the (anti-)BRST symmetry transformations (39) and (40) on the constrained 
super world-line defined by the CF-type 
restriction ($b + \bar b + 2\;\beta\;\bar\beta = 0$). Using the standard techniques of BRST-formalism, one can write the following    
\begin{eqnarray}
L^{(0)}_b = L_0 + s_b\;s_{ab} \;\Bigl[\frac{i}{2}\; e^2 + c\;\bar c\Bigr],
\end{eqnarray}
\begin{eqnarray}
L^{(0)}_{\bar b} = L_0 - s_{ab}\; s_b \;\Bigl[\frac{i}{2}\; e^2 + c\;\bar c\Bigr],
\end{eqnarray}
where $L_0$ is the starting Lagrangian (1) and $s_{(a)b}$ are the nilpotent
and absolutely anticommuting (anti-)BRST transformations (39) and (40). It is straightforward 
to check that
\begin{eqnarray}
s_b\;s_{ab} \;\Bigl[\frac{i}{2}\; e^2 + c\;\bar c\Bigr] &=& - i\; \dot{\bar c}\;(\dot c + 2\;\beta\;\chi) + 2\;i\;\bar\beta \;\dot c\;\chi
- \dot b\;e - 2\;e\;(\gamma\;\chi + \bar\beta\;\dot\beta) \nonumber\\&+& 2\;\beta\;\gamma\;\bar c - b\;\bar b + \bar\beta^2\;\beta^2 + 
2\;\bar\beta\; c\;\gamma,
\end{eqnarray}
\begin{eqnarray}
- s_{ab}\;s_b \;\Bigl[\frac{i}{2}\; e^2 + c\;\bar c\Bigr] &=& - i\; \dot{\bar c}\;(\dot c 
+ 2\;\beta\;\chi) + 2\;i\;\bar\beta \;\dot c\;\chi
+ \dot {\bar b}\;e - 2\;e\;(\gamma\;\chi - \beta\;\dot{\bar\beta}) \nonumber\\
&+& 2\;\beta\;\gamma\;\bar c - b\;\bar b + \bar\beta^2\;\beta^2 + 
2\;\bar\beta\; c\;\gamma,
\end{eqnarray}
which are nothing but the gauge-fixing and Faddeev-Popov ghost terms for the present theory (being described within the framework of BRST formalism).

Throwing away the total derivative terms and using the CF-restriction 
($b + \bar b + 2\;\beta\;\bar\beta = 0$), we obtain the following form of the coupled (but equivalent) Lagrangians 
\begin{eqnarray}
L^{(0)}_b &=& L_0 + b \;\dot e + b\;(b + 2\;\beta\;\bar\beta) - i\; \dot{\bar c}\;(\dot c 
+ 2\;\beta\;\chi) + 2\;i\;\bar\beta \;\dot c\;\chi
\nonumber\\&-& 2\;e\;(\gamma\;\chi + \bar\beta\;\dot\beta) + 2\;\beta\;\gamma\;\bar c 
+ \bar\beta^2\;\beta^2 + 2\;\bar\beta\; c\;\gamma,
\end{eqnarray}
\begin{eqnarray}
L^{(0)}_{\bar b} &=& L_0  - \bar b \;\dot e + \bar b\;(\bar b + 2\;\bar\beta\;\beta)
- i\; \dot{\bar c}\;(\dot c + 2\;\beta\;\chi) + 2\;i\;\bar\beta \;\dot c\;\chi
\nonumber\\ &-& 2\;e\;(\gamma\;\chi - \beta\;\dot{\bar\beta}) + 2\;\beta\;\gamma\;\bar c 
+ \bar\beta^2\;\beta^2 + 2\;\bar\beta\; c\;\gamma.
\end{eqnarray}
We note that we have expressed ($b\;\bar b$), present in equations (49) and (50), 
in two different ways by using the CF-type restriction ($b + \bar b + 2\;\beta\;\bar\beta = 0$).
We can check explicitly that, the following {\it perfect} symmetry transformations emerge, namely;
\begin{eqnarray}
s_b\; L^{(0)}_b = \frac{d}{d\tau}\;\Bigl[\;\frac{1}{2}\;c\; p^2 + \frac{\beta}{2}\;(p \cdot\psi)  
+ b\;(\dot c + 2\; \beta\; \chi) \Bigr],
\end{eqnarray}
\begin{eqnarray}
s_{ab}\; L^{(0)}_{\bar b} = \frac{d}{d\tau}\;\Bigl[\;\frac{1}{2}\;\bar c\; p^2 + \frac{\bar\beta}{2}\;(\;p \cdot\psi)  
- \bar b\;(\dot {\bar c} + 2\; \bar\beta\; \chi)\Bigr].
\end{eqnarray}
As a consequence, the action integrals ($S_1 = \int d\tau\; L^{(0)}_b,\; S_2 = \int d\tau\;L^{(0)}_{\bar b}$)
remain invariant under the (anti-)BRST transformations (39) and (40). There are other ways
of expressing (47) and (48) as we have done in our Appendices A and B. However, we choose the
forms (51) and (52) because, at least, these respect {\it perfect} symmetries like (53) and (54).

The Lagrangians (51) and (52) are {\it equivalent} on the  constrained super world-line, defined by 
the CF-type condition (30). This can be corroborated by the following observations
\begin{eqnarray}
s_{ab}\; L^{(0)}_ b &=& \frac{d}{d\tau}\;\Bigl[\;\frac{1}{2}\;\bar c\; p^2 + \frac{1}{2}\; \bar\beta\;(\;p \cdot\psi)
+ 2\;i\;e\;\bar\beta\;\gamma  + b\;(\dot {\bar c} + 2\; \bar\beta\; \chi)\Bigr] \nonumber\\
&-& (\dot{\bar c} + 2\;\bar\beta\;\chi)\;\frac{d}{d\tau}\;\Bigl[b + \bar b + 2\;\beta\;\bar\beta \Bigr] + (2\;i\;\bar\beta\;\gamma)\;(b + \bar b + 2\;\beta\;\bar\beta),
\end{eqnarray}
\begin{eqnarray}
s_b\; L^{(0)}_{\bar b} &=& \frac{d}{d\tau}\;\Bigl[\;\frac{1}{2}\;c\; p^2 + \frac{1}{2}\;\beta\;(\;p \cdot\psi) + 2\;i\;e\;\beta\;\gamma   
- \bar b\;(\dot c + 2\;\beta\; \chi)\Bigr] \nonumber\\
&+& (\dot c + 2\;\beta\;\chi)\;\frac{d}{d\tau}\;\Bigl[b + \bar b + 2\;\beta\;\bar\beta \Bigr] - (2\;i\;\beta\;\gamma)\;(b + \bar b + 2\;\beta\;\bar\beta),
\end{eqnarray}
which demonstrate the equivalence of $L^{(0)}_b$ and $L^{(0)}_{\bar b}$ [as far as the
(anti-)BRST symmetry transformations (39) and (40) are concerned] on the constrained super world-line
defined by the CF-type restriction (30). In other words, both $L^{(0)}_b$ and $L^{(0)}_{\bar b}$
respect the off-shell nilpotent (anti-)BRST symmetry transformations (39) and (40) if we confine ourselves to the
constrained super world-line [defined by the CF-condition (30)] embedded in the D-dimensional target 
Minkowaskian flat spacetime manifold. We obtain, from the above Lagrangians (51) and (52), the following
very useful relationships, namely;    
\begin{eqnarray}
b = - \frac{1}{2}\; \dot e - \beta\;\bar\beta, \qquad\qquad \bar b =  \frac{1}{2}\; \dot e - \beta\;\bar\beta,    
\end{eqnarray}  
as the Euler-Lagrange equations of motion which, ultimately, lead to the derivation of the CF-type condition 
($b + \bar b + 2\;\beta\;\bar\beta = 0$) in a straightforward manner.

The (anti-)BRST invariance (54) and (53) of $L_{\bar b}^{(0)} $ and $L_{b}^{(0)}$ can be captured within the framework
of superfield formalism. Towards this goal in mind, first of all, we check that 
the super Lagrangian $\tilde L_0$, corresponding to the starting Lagrangian $L_0$, 
can be written in the following form in terms of the appropriate superfields:
\begin{eqnarray}
{\tilde L}_0 = P^{(g)}\cdot \dot X^{(sg)} - \frac{1}{2}\;E^{(sh)}\; (P^{(g)})^2 
+ \frac{i}{2}\; \Psi^{(sg)}\cdot\dot \Psi^{(sg)}  
+ i \;K^{(h)}\; (P^{(g)}\cdot \Psi^{(sg)}) \equiv L_0, 
\end{eqnarray}
where the super-expansions of the relevant superfields, after the application of (S)GIR and (SUSY-)HC, are quoted in 
(38). It is obvious from (58) that the l.h.s. is independent of $\theta$ and $\bar\theta$ variables (as ${\tilde L}_0 = L_0$).
As a consequence, we have the following relationship
\begin{eqnarray}
\lim_{\theta \to 0}  \;\frac {\partial}{\partial{\bar\theta}} 
\;{\tilde L}_0 = 0, \quad \qquad \quad \;\; \lim_{\bar\theta \to 0}  \;\frac {\partial}{\partial{\theta}}
\;{\tilde L}_0 = 0,
\end{eqnarray} 
which, in the ordinary 1D spacetime, imply that the starting Lagrangian $L_0$ remains invariant
under the (anti-)BRST transformations  (39) and (40).

In view of the above observation, it is straightforward  to express the Lagrangians (47) and (48) in terms of the superfields, obtained 
after the application of (anti-)BRST invariant (SUSY-)HC and (S)GIR, as follows 
\begin{eqnarray}
&&{\tilde L}^{(0)}_b = {\tilde L}_0 + \frac {\partial}{\partial{\bar\theta}}\; \frac {\partial}{\partial{\theta}}
\;\Bigl[\frac{i}{2} \;E^{(sh)} \; E^{(sh)}\; + \;\;F^{(sh)} \; \bar F^{(sh)}\;\Bigr], \nonumber\\
&&{\tilde L}^{(0)}_{\bar b} = {\tilde L}_0 - \frac {\partial}{\partial{\theta}}\; \frac {\partial}{\partial{\bar\theta}}
\;\Bigl[\frac{i}{2} \;E^{(sh)} \; E^{(sh)}\; + \;\;F^{(sh)} \; \bar F^{(sh)}\;\Bigr],
\end{eqnarray}
where the expansions for the superfields are quoted in (38). The (anti-)BRST invariance and equivalence 
of the Lagrangians $L^{(0)}_b$ and $L^{(0)}_{\bar b}$ can be captured, in a very simple manner, within the framework of 
superfield formalism as illustrated below
\begin{eqnarray}
\lim_{\theta \to 0}  \;\frac {\partial}{\partial{\bar\theta}} 
\;{\tilde L}_{(b, \bar b)} = 0, \quad \qquad \quad \;\; \lim_{\bar\theta \to 0}  \;\frac {\partial}{\partial{\theta}}
\;{\tilde L}_{(b, \bar b)} = 0,
\end{eqnarray}
where the nilpotency property ($\partial^2_\theta = 0, \; \partial^2_{\bar\theta} = 0$) of the Grassmannian derivative 
($\partial_\theta, \partial_{\bar\theta}$) and the their absolute anticommutativity
($ \partial_\theta\; \partial_{\bar\theta} + \partial_{\bar\theta}\; \partial_\theta = 0$) play a very decisive role.
Furthermore, the (anti-)BRST invariance of the starting Lagrangian $L_0$, in the language of the superfield formalism, has also been taken into consideration in the proof of (61).

We close this section with a few remarks. First and foremost, the Lagrangian,
proposed in [11] for the  spinning relativistic particle, does not respect 
the (anti-)BRST symmetries {\it together}. Second, the coupled (but equivalent) 
Lagrangians of (47) and (48) are the {\it correct} Lagrangians for the supersymmetric 
system of a spinning relativistic particle  which respect both the above nilpotent 
symmetries on a constrained super world-line [where the CF-type restriction 
$b + \bar b + 2\;\beta\;\bar\beta = 0$
is satisfied]. Finally, it is the augmented version of BT superfield formalism [8-10]
that plays a key role in the derivation of the proper (anti-)BRST symmetries 
and the (anti-)BRST invariant CF-type restriction. The latter, we claim, is the hallmark 
of a supersymmetric gauge theory within the framework of BRST formalism.

\section{Conclusions}

One of the key results of our present investigation is the derivation of  supersymmetric version of HC (SUSY-HC)
in equation (22) which enables us to derive the off-shell nilpotent and absolutely anticommuting
(anti-)BRST transformations for the variables $e(\tau), \;c(\tau),\; \bar c(\tau)$ and $x_\mu(\tau)$ within
the framework of supersymmetric version of the augmented BT superfield formalism. The beauty of this SUSY-HC (22)
lies in the fact that it utilizes the results of HC [cf. (7), (23)] in a meaningful manner
and produces [cf. (30)] the (anti-)BRST invariant CF-type restriction\footnote{It is to be noted that, for the first-time, the celebrated Curci-Ferrari condition [15]
appeared in the description of the 4D non-Abelian 1-form gauge theory within the framework of 
BRST formalism.}. As a consequence, the
(anti-)BRST symmetry transformations of $\chi(\tau),\; \beta(\tau), \;\bar\beta(\tau)$ and $\psi_\mu(\tau)$ 
(derived from the application of HC) turn out to be 
consistent and complementary to such kind of transformations for $e(\tau),\; c(\tau),\; \bar c(\tau)$ and $x_\mu(\tau)$
(derived from the power and beauty of SUSY-HC).
The rest of the (anti-)BRST symmetry transformations (e.g. for the auxiliary variables) are obtained from the requirements of 
off-shell nilpotency and absolute anticommutativity of the (anti-)BRST transformations.

The supersymmetric system of a massless spinning relativistic particle  is a physically very interesting system 
where we have applied the augmented version of BT-superfield  formalism {\it for the first-time}
and derived the proper (i.e. off-shell nilpotent and absolutely anticommuting) anti-BRST symmetry transformations. Besides 
HC and SUSY-HC, in our investigation, we have shown that the (super)gauge invariant quantities also play very important 
roles. This is why, there are different varieties of superscripts in expansions (38). The central observation of our 
present work is the fact that all the proper nilpotent (anti-) BRST transformations, derived from HC, SUSY-HC, GIR and SGIR,
are consistent with one-another. We would like to mention that we have {\it never} used so many conditions on
the superfields in our earlier study of non-supersymmetric $p$-form gauge theories.

It is very interesting to point out that the HC yields the proper (anti-)BRST symmetry transformations for the
super (fermionic) gauge variable $\chi (\tau)$ and its associated bosonic (anti-)ghost variables $(\bar\beta)\beta$.
To obtain the (anti-)BRST symmetry  transformations for the fermionic variable $\psi_\mu (\tau)$, however,
we have been theoretically compelled to use the equation of motion $\dot \psi_\mu = \chi\; p_\mu$ which
is a supergauge invariant quantity. In exactly similar fashion, the SUSY-HC [cf. (22)] yields the proper
(anti-)BRST symmetry transformations for the bosonic gauge variable $ e (\tau)$ and its associated
fermionic (anti-)ghost variables $(\bar c) c$. However, to obtain the proper (anti-)BRST symmetry
transformations for the target space (bosonic) variable $x_\mu (\tau)$, we have  invoked another equation
of motion $\dot x_\mu = e p_\mu - i \chi \psi_\mu$ which is {\it also} a supergauge invariant quantity. It is
precisely, because of these observations, that we have christened our superfield formalism as the supersymmetric
version of the augmented BT superfield formalism where (super)gauge invariance plays a decisive role.

We would like to point out that, in [11], only the nilpotent
BRST symmetries for the free spinning relativistic particle
have been discussed. However, the corresponding proper anti-BRST symmetries have been left untouched. In our recent couple 
of papers [16,17], we have established that the existence of anti-BRST symmetry transformations
is sacrosanct in the context of BRST description of any arbitrary $p$-form ($p = 1, 2, 3,...$) 
gauge theories as it is crucially
connected with the existence of CF-type of restrictions which owe their origin to the geometrical object called gerbes.
In fact, we have claimed that, given a local gauge symmetry transformation, the holy grails of the BRST formalism 
allow us to have both BRST as well as anti-BRST symmetry transformations {\it together} in a gauge theory. The decisive
feature of a gauge theory, within the framework of BRST formalism, is the existence of the CF type restriction
(which allows totally independent existence of BRST as well as anti-BRST symmetry transformations). For the simple case of Abelian
1-form gauge theory, the CF type restriction is {\it trivial}. However, for the rest of the $p$-form gauge theories,
the CF type restriction is always {\it non-trivial} (see, e.g. [16,17]).

In our present investigation, we have applied our superfield formalism to a simple supersymmetric
system of a free (massless as well as massive) spinning relativistic particle\footnote{We have
applied the theoretical arsenal of superfield approach to BRST
formalism in the case of massless spinning relativistic particle in great detail. However, we
have concisely mentioned the application of superfield formalism to the derivation of
(anti-)BRST symmetries for the massive spinning relativistic particle in our Appendix C. The details
of the superfield technique can be worked out for the latter system, too, on exactly same lines as 
that of the massless spinning particle.}. 
In the future, we plan to extend this work to the description of some
other physically interesting supersymmetric models and derive various conserved charges (corresponding to   
various continuous symmetries) of the theory which would include the conserved (super)charges, (anti-)BRST charges, 
ghost charges, etc., and we envisage to obtain the underlying algebra and look into its relevance to the algebra 
of some (super)Lie groups. Furthermore, we hope to apply the supersymmetric version
of the augmented BT superfield formalism to more physically  realistic SUSY models of phenomenological
interest as far as the $p$-form gauge theories are concerned. These are some of the issues that are presently under investigation and our results would
be reported in our future publications [18].\\

\noindent
{\bf Acknowledgements:} Two of us (AS and SK) remain grateful to UGC, Govt. of India, New Delhi, 
for financial support under RFSMS  and RGNF schemes. The present investigation has
been carried out under the auspices of the above fellowships. We remain grateful to our esteemed referees for 
their enlightening and stimulating comments. \\

\vspace {1 cm}

\hspace{6cm}{\bf {\large Appendix A}}

\vspace{0.8 cm}

\noindent
We derive here two Lagrangians, by exploiting (39) and (40), which respect anti-BRST and BRST symmetry 
transformations but they are {\it not} equivalent even on the hyper super 
world-line defined by the
CF-type restriction. As a consequence, they are {\it not} as interesting as (51) and (52).
Using the standard techniques of BRST formalism, we can obtain the Lagrangian that respects only the  BRST transformations (40). Such a Lagrangian is: 
\[L^{(1)}_b = L_0 + s_b \Bigg[-i \;\bar c \left(\dot e\; + \;\frac{b}{2}\right)
\;- \;i \;\bar\beta\; \dot \chi \Bigg],\eqno{(A.1)} \]
where $L_0$ is our starting Lagrangian (1). We have to exploit the BRST transformations 
(40) to obtain the following explicit form of the BRST invariant Lagrangian
\[ L^{(1)}_b = L_0 + b \;\dot e + \frac{1}{2}\;b^2 - i \;\dot {\bar c}\; (\dot c + 2\; \beta\; \chi) 
+ \gamma \; \dot\chi - \dot {\bar\beta} \dot \beta, \eqno{(A.2)} \]
which remains  quasi-invariant under $s_b$ because
\[ s_b\; L^{(1)}_b = \frac{d}{d\tau}\;\Bigl[\;\frac{1}{2}\;c\; p^2 
+ \frac{1}{2}\; \beta \;(p \cdot\psi)  + b\;(\dot c + 2\; \beta\; \chi)
- i\;\gamma\; \dot\beta\; \Bigr ]. \eqno{(A.3)} \]
As a consequence, the action integral $S = \int d\tau L^{(1)}_b$ remains invariant for the physically well-defined variables 
that vanish off at infinity due to Gauss's divergence theorem.

In exactly similar fashion, we can derive the anti-BRST invariant Lagrangian by exploiting the anti-BRST transformations (39).
This Lagrangian ($L^{(1)}_{\bar b}$) can be written as 
\[ L^{(1)}_{\bar b} = L_0 + s_{ab} \; \Bigg[i \; c \;\left(\dot e\; + \;\frac{\bar b}{2} \right)\;
+ \;i \;\beta\; \dot \chi \Bigg], \eqno{(A.4)} \]
where $L_0$ is our starting Lagrangian (1). Exploiting the anti-BRST symmetry transformations (39), we
obtain the following explicit form of the anti-BRST invariant Lagrangian
\[ L^{(1)}_{\bar b} = L_0 - \bar b \;\dot e - \frac{1}{2}\;{\bar b}^2 + i \;\dot c \; (\dot {\bar c} 
+ 2\; \bar\beta\; \chi) + \gamma \; \dot\chi + \dot {\bar\beta} \dot \beta. \eqno{(A.5)} \]
The anti-BRST invariance of $L^{(1)}_{\bar b}$ can be checked by using the transformations (39) as we note that $L^{(1)}_{\bar b}$
transforms to a total derivative as given below 
\[ s_{ab}\; L^{(1)}_{\bar b} = \frac{d}{d\tau}\;\Bigl[\;\frac{1}{2}\;\bar c\; p^2 
+ \frac{1}{2}\; \bar\beta \;(p \cdot\psi)  
- \bar b\;(\dot {\bar c} + 2\; \bar\beta\; \chi) 
- i\; \gamma \;\dot{\bar\beta}\; \Bigr]. \eqno{(A.6)} \]
As a consequence, the action integral $S = \int d\tau L^{(1)}_{\bar b}$ remains invariant 
for the physically well-defined variables of the theory which fall-off rapidly at infinity.

It is evident from (A.2)
 and (A.5) that both the Lagrangians are coupled because we have already derived 
the CF-type restriction $b +\bar b + 2\; \beta \; \bar\beta= 0$ which relates 
the Nakanishi-Lautrup type of variables $b$ and $\bar b$ through the bosonic 
(anti-)ghost variables $(\bar\beta)\beta$. We have already captured the CF-type
restriction within the framework of superfield formalism. Now we demonstrate that 
the (anti-)BRST invariance [cf. (A.3), (A.6)] of Lagrangians (A.2) and (A.5) 
(and the Lagrangians themselves) can also be incorporated in the language of the
superfield formalism. We note that the Lagrangians (A.2) and (A.5) can be expressed in the language of
superfields [obtained after (SUSY-)HC and (S)GIR], as 
\[ {\tilde L}^{(1)}_b = {\tilde L}_0 + \lim_{\theta \to 0}  \;\frac {\partial}{\partial{\bar\theta}}
\left[-i \;\bar F^{(sh)} \left(\dot E^{(sh)}\; + \;\frac{b (\tau)}{2}\right) - \;i \;\bar {\cal B}^{(h)}\; \dot K^{(h)} \right], \nonumber\] 
\[ {\tilde L}^{(1)}_{\bar b} = {\tilde L}_0 + \lim_{\bar\theta \to 0}  \;\frac {\partial}{\partial{\theta}}
\left [i \;F^{(sh)} \left(\dot E^{(sh)}\; + \;\frac{\bar b (\tau )}{2} \right)
+ \;i \;{\cal B}^{(h)}\; \dot K^{(h)} \right],\eqno{(A.7)} \]
where the super-expansions of all the superfields (with various superscripts) are given in (38).
In view of our observation in (59), it is elementary to show that 
\[ \lim_{\theta \to 0}  \;\frac {\partial}{\partial{\bar\theta}} 
\;{\tilde L}^{(1)}_b = 0, \quad \qquad \quad \;\; \lim_{\bar\theta \to 0}  \;\frac {\partial}{\partial{\theta}}
\;{\tilde L}^{(1)}_{\bar b} = 0, \eqno{(A.8)} \]
which, because of the mappings (43), establish the (anti-)BRST invariance of $(L_{\bar b})L_b$
in the ordinary 1D spacetime [cf. (A.3),(A.6)]. We mention here that it is the (anti-)BRST
invariance of $L_0$ [cf. (59)] and the nilpotency ($s^2_{(a)b} = 0$) of (anti-)BRST symmetry transformations 
(i.e. $\partial^2_\theta = \partial^2_{\bar\theta} = 0 $) that have played the key roles in the proof of (A.8).

Before we wrap up this Appendix, we note that, even though, the Lagrangians (A.2) and (A.5) respect the (anti-)BRST
symmetries [cf. (A.6) , (A.3)], they are not {\it equivalent} even on the super world-line defined by CF-type
condition ($b + \bar b +2\;\beta\;\bar\beta = 0 $). This can be checked in the following manner
by taking the help of symmetry properties, namely;
\[ s_b\; L^{(1)}_{\bar b} = \frac{d}{d\tau} \left[\frac {1}{2}\;c\; p^2 
+ \frac{1}{2}\;\beta\;\left(p \cdot \psi \right) - \bar b\; \left(\dot {c} 
+ 2\;\beta\;\chi \right) - i\; \gamma\; \dot{\beta} \right] + \left (\dot {b} + \dot {\bar b} 
+2\;\bar\beta\;\dot{\beta} \right) \dot{c}
\nonumber\] \[ +2\;i\;\beta\;\gamma \left(\bar b + \dot {e} \right) + 2\;\left( i\; \dot{\gamma} 
+ \beta\; \dot{\bar c} \right)\dot{\beta} + 2\left[ \dot {\bar b} \;\beta 
+ 2\;\beta\;\dot{\beta}\; \bar\beta + \dot{c}\; \gamma\right] \chi, \eqno{(A.9)} \]
\[ s_{ab} \;L^{(1)}_b = \frac{d}{d\tau} \left[\frac {1}{2}\;\bar c\; p^2 
+ \frac{1}{2}\;\bar\beta\;\left(p \cdot \psi \right) + b\; \left(\dot {\bar c} 
+ 2\;\bar\beta\;\chi \right) - i\; \gamma\; \dot{\bar\beta} \right] -\dot{\bar c} \left (\dot {b} + \dot{\bar b} 
+2\;\beta\;\dot{\bar\beta}\right) 
\nonumber\] \[ +2\;i\;\bar\beta\;\gamma \left(b + \dot {e} \right) + 2\;\left( i\; \dot{\gamma} 
- \bar\beta\; \dot{c} \right)\dot{\bar\beta} + 2\left( \dot{\bar c}\; \gamma - \dot {b} \;\bar\beta 
- 2\;\beta\;\bar\beta\;\dot{\bar\beta}\right) \chi. \eqno{(A.10)} \]
Thus, we lay emphasis on the fact that the Lagrangians $L^{(1)}_b$ and $L^{(1)}_{\bar b}$ (even though endowed with some interesting properties) are {\it not} equivalent because even if we use the equations of motion and/or
the celebrated CF-type restriction ($b + \bar b +2\;\beta\;\bar\beta = 0$), we do {\it not} find the 
precise (anti-)BRST invariance of ($L^{(1)}_{\bar b}$)$L^{(1)}_b$ [as is evident from equations (A.9) and A.10)]. Furthermore, the other drawback of the Lagrangians (A.3) and (A.5) is the fact that we are unable to obtain the CF-type restriction as an off-shoot 
from the Euler-Lagrange equations of motion (derived from  the Lagrangians $L^{(1)}_b$ and $L^{(1)}_{\bar b}$).

\vspace {1 cm}

\hspace{6cm}{\bf {\large Appendix B}}

\vspace{0.8 cm}

\noindent
We show here that the Lagrangians  
(47) and (48) (cf. Sec. 6) can be expressed in a coupled 
form that are different from the ones given in (51) and (52). As we have
seen, the latter forms have their own merits. However, the former ones
are more symmetrical (in some sense) because the Nakanishi-Lautrup type of auxiliary
variables $b$ and $\bar b$ appear in these Lagrangians in a symmetrical manner. 
Such coupled Lagrangians, which are more symmetrical in form, are as follows
\[ L^{(2)}_b = L_0 + b\; \dot{e} + \frac {1}{2} \left(b^2 + {\bar b}^2 \right) 
-i\; \dot{\bar c} \left(\dot {c} + 2\; \beta\;\chi \right) +2\;i\;\bar\beta \;\dot{c}\;\chi \nonumber\]
\[- 2\;e\;( \gamma\;\chi +\bar\beta\; \dot{\beta})
- 2\;\bar c\;\beta\; \gamma - \bar\beta^2\;\beta^2 + 2\; \bar\beta\;c\;\gamma, \eqno{(B.1)}\]
\[ L^{(2)}_{\bar b} = L_0 - {\bar b}\; \dot{e} + \frac {1}{2} \left(b^2 + {\bar b}^2 \right) -i\; \dot{\bar c} \left(\dot {c} + 2\; \beta\;\chi \right) +2\;i\;\bar\beta \;\dot{c}\;\chi \nonumber\]
\[- 2\;e\;( \gamma\;\chi -\beta\; \dot{\bar\beta})
- 2\;\bar c\;\beta\; \gamma - \bar\beta^2\;\beta^2 + 2\; \bar\beta\;c\;\gamma, \eqno{(B.2)}\]
where we have used the following standard relationship
\[b\;\bar b = \frac{1}{4} \left(b +\bar b \right)^2 - \frac{1}{4} \left(b - \bar b \right)^2 \equiv 2\;\bar\beta^2\;\beta^2 - \frac{1}{2} \left(b^2 + {\bar b}^2 \right), \eqno{(B.3)}\]
due to the CF-type restriction $b + {\bar b} = -2\;\beta\;\bar\beta$. The coupled
Lagrangians (B.1) and (B.2) are {\it equal} on the constraint world-line defined by
$b + \bar b + 2\;\beta\;\bar\beta = 0$.
In other words, the terms, that differ between (B.1) and (B.2), are primarily equal 
due to the CF-type restriction. We note that the CF-type 
equation can be derived from the following equations of motion
\[\dot {b} = -\frac{1}{2}\; p^2 - 2\;\gamma\;\chi - 2\; \bar\beta\;\dot{\beta} \nonumber\]
\[ \dot {\bar b} = +\frac{1}{2}\; p^2 + 2\;\gamma\;\chi - 2\; \beta\;\dot{\bar\beta}, \eqno{(B.4)}\]
that emerge from the Lagrangian $L^{(2)}_b$ and $L^{(2)}_{\bar b}$.

Now we discuss the symmetry properties of (B.1) and (B.2) under the nilpotent and
absolutely anticommuting (anti-)BRST symmetry transformations (39) and (40). 
It can be easily checked that, under (40), 
the Lagrangian $L_b$ transforms as follows
\[ s_b\; L^{(2)}_b = \frac{d}{d\tau} \left[\frac {1}{2}\;c\; p^2 
+ \frac{1}{2}\;\beta\;\left(p \cdot \psi \right) + b\; \left(\dot {c} 
+ 2\;\beta\;\chi \right)  \right] - 2\;i\;\beta\;\gamma \left(b 
+ \bar b +2\;\beta\;\bar\beta \right).\eqno{(B.5)}\]
Similarly, under the off-shell nilpotent anti-BRST transformations (39), the other 
Lagrangian $L^{(2)}_{\bar b}$ (of the present physical system) transforms to a total derivative as
\[ s_{ab}\; L^{(2)}_{\bar b} = \frac{d}{d\tau} \left[\frac {1}{2}\;\bar c\; p^2 
+ \frac{1}{2}\;\bar\beta\;\left(p \cdot \psi \right) - \bar b\; \left(\dot {\bar c} 
+ 2\;\bar\beta\;\chi \right) \right] + 2\;i\;\bar\beta\;\gamma \left(b + \bar b 
+2\;\beta\;\bar\beta \right).\eqno{(B.6)}\]
Thus, we note that ($L^{(2)}_{\bar b}$)$L^{(2)}_b$ respect (anti-)BRST 
symmetry transformations $s_{(a)b}$ {\it only} when the CF-type condition 
($b + \bar b +2\; \beta\;\bar\beta = 0$) is satisfied. In other words,
the off-shell nilpotent (anti-)BRST transformations (39) and (40) are symmetry transformations
for the Lagrangians $L^{(2)}_{\bar b}$  and $L^{(2)}_b$ only on a constrained super world-line, 
defined by the CF-type equation ($b + \bar b +2\; \beta\;\bar\beta = 0$), which is 
embedded in the D-dimensional target spacetime Minkowaskian flat manifold. It can be easily noted 
that the symmetry transformations in (B.5) and (B.6) are {\it not} like the transformations in (53)
and (54). Hence, the former are {\it not} perfect symmetry transformations as are the latter set [cf. (53),(54)].

We close this Appendix with the following remarks. First, the (anti-)BRST 
transformations (39) and (40) are {\it not perfect} symmetry transformations 
for the Lagrangians $L^{(2)}_{\bar b}$ and $L^{(2)}_b$. Rather, these are the 
symmetry transformations only under the validity of the CF-type restriction. 
Second, the Lagrangians $L^{(2)}_b$ and $L^{(2)}_{\bar b}$, it can be checked 
explicitly, transform exactly same as the Lagrangians $L^{(0)}_b$ and $L^{(0)}_{\bar b}$ 
of Sec. 6 under the transformations $s_{ab}$ and $s_b$, respectively [cf. (55), (56)]. 
Third, the (anti-)BRST invariance of Lagrangians $L^{(2)}_b$ and $L^{(2)}_{\bar b}$ can be captured 
within the framework of superfield formalism, in exactly same manner, as we have accomplished this goal in Sec. 6.
Fourth, as far as the {\it symmetrical} form is concerned, the coupled Lagrangians (B.1) and (B.2) are more beautiful
than (51) and (52). However, from the point of view of {\it perfect} symmetry, the Lagrangians (51) and (52) are more
appropriate [cf. (53),(54)].

\vspace {1 cm}

\hspace{6cm}{\bf {\large Appendix C}}

\vspace{0.8 cm}

\noindent
For the present paper to be complete and self-contained, we discuss here briefly the superfield approach to BRST
analysis of the massive spinning relativistic particle.
The analogue of the Lagrangian (1), for the free massive relativistic spinning relativistic particle, is [11]
\[
L_1 = p_\mu\; \dot x^\mu - \frac{e}{2}\; (p^2 + m^2) + \frac{i}{2}\; (\psi_\mu \;\dot \psi^\mu - \psi_5\; \dot\psi_5)  
+ i \;\chi\; (p_\mu\; \psi^\mu - \psi_5 \; m),\eqno{(C.1)}\]
where the $\tau$-independent mass parameter $m$ (which happens to be the analogue 
of the cosmological constant term) has been
introduced by invoking a Lorentz scalar fermionic (i.e. $\chi\; \psi_5 + \psi_5 \;\chi = 0, \psi_5 \;\psi_\mu
+ \psi_\mu \;\psi_5$, etc.)
variable $\psi_5$ (with $\psi^2_5 = -1$).

We note that the analogue of the combined (super)gauge transformations (3), in our present case
of a free massive spinning particle, are
\[\delta\; x_\mu = \xi\; p_\mu  + \kappa\; \psi_\mu, \qquad \delta\; p_\mu = 0,
\qquad \delta \;\psi_\mu = i\;\kappa\; p_\mu,\nonumber\]
\[\delta\; \chi = i\; \dot \kappa, \;\;\qquad \delta \;\psi_5 = i\;\kappa\;m, 
\;\;\qquad \delta \;e = \dot \xi + 2\;\kappa\; \chi.\eqno{(C.2)}\]
Under the above transformations, the Lagrangian $L_1$ transforms as
\[ \delta L_1 = \frac {d}{d\tau} \;\Big [ \frac {\xi}{2} \;(p^2 + m^2)
+ \frac {\kappa}{2} \;(p \cdot \psi + m\; \psi_5)\Big]. \eqno{(C.3)}\]
As a consequence, the transformations (C.2) are the symmetry transformations for
the action integral $S =\int d\tau\; L_1$ which remains invariant for the physically well-defined
dynamical variables of the theory that vanish off at infinity.

The proper (i.e. off-shell nilpotent and absolutely anticommuting) (anti-)BRST transformations, 
corresponding to the (super)gauge symmetry transformations (C.2), can be obtained by our 
geometrical superfield formalism. In particular, we can derive the nilpotent (anti-)BRST symmetry 
transformations of $\psi_5$ by using the Euler-Lagrange equation of motion corresponding to 
the dynamical variable $\psi_5$ (which is $\dot\psi_5 = \chi\; m$). Furthermore, it can be checked
that $\dot\psi_5 = \chi\; m$ is a super-gauge invariant quantity. 
Therefore, the above equation (according to the basic tenets of the augmented version of BT superfield
formalism [8-10,14]) can be written in terms of the superfields as 
  \[
\dot\Psi_5(\tau, \theta, \bar\theta) = K^{(h)}(\tau, \theta, \bar\theta) \; m \;\;\Rightarrow\;\;
\dot\Psi_5(\tau, \theta, \bar\theta) - K^{(h)}(\tau, \theta, \bar\theta)\; m = 0,
\eqno{(C.4)}\] 
where the expansion for the superfield $\Psi_5 (x, \theta, \bar\theta)$ is taken to be
\[
\Psi_5(\tau, \theta, \bar\theta) = \psi_5(\tau) + \theta\; \bar B_5(\tau) + \bar\theta\; B_5(\tau) + \theta\;\bar\theta\;f_5(\tau). \eqno{(C.5)}\]  
In the above super expansion, it is elementary to note that $\psi_5 (\tau)$ and $f_5 (\tau)$ are fermionic
in nature as against the bosonic nature of $B_5 (\tau)$ and $\bar B_5 (\tau)$. The secondary variables
[$B_5 (\tau), \bar B_5 (\tau), f_5 (\tau)$] are to be determined in terms of the basic and auxiliary
variables of the 1D (anti-)BRST invariant theory by exploiting the restriction (C.4). Furthermore,
to obtain the correct results, we have to use the expansions for
$K^{(h)}(\tau, \theta, \bar\theta)$ from equation (38). To be precise, we  
equate the coefficients of  $\theta, \bar\theta$ and $\theta\;\bar\theta$ to zero  that emerge
from the super-gauge invariant restriction (C.4). This requirement, in fact, leads to
\[
{\dot{\bar B}}_5 = i\;\dot{\bar\beta}\; m \;\;\;\Longrightarrow \;\;\;\bar B_5 = i\;\bar\beta\; m, \nonumber\]
\[ {\dot B}_5 = i\;\dot\beta\; m \;\;\;\Longrightarrow \;\;\; B_5 = i\;\beta\; m, \nonumber\]
\[{\dot f}_5 = - \dot\gamma\; m \;\;\;\Longrightarrow \;\;\; f_5 = - \gamma\; m. \eqno(C.6) \]
As a consequence, the superfield expansion of $\Psi_5(\tau, \theta, \bar\theta)$ turns out to be
\[
\Psi^{(sg)}_5(\tau, \theta, \bar\theta) = \psi_5 + \theta\; (i\;\bar\beta\; m) + \bar\theta\;(i\;\beta\; m) + \theta\;\bar\theta\;(- \gamma\; m).\eqno(C.7)\]
In the language of the (anti-)BRST symmetry transformations (39) and (40), the above expansion can
be re-expressed in the following manner, namely;
\[ \Psi^{(sg)}_5(\tau, \theta, \bar\theta)
\equiv \psi_5 + \theta\; (s_{ab}\;\psi_5) + \bar\theta\;(s_b\;\psi_5) + \theta\;\bar\theta\;(s_b\;s_{ab}\;\psi_5).
\eqno(C.8)\]   
From the above, it is clear that
we obtain the proper (anti-)BRST symmetry transformations for the dynamical variable $\psi_5 (\tau)$ as follows
\[
s_b\;\psi_5 = i\;\beta\; m, \quad\qquad s_{ab}\;\psi_5 = i\;\bar\beta\; m,\quad\qquad s_b\;s_{ab}\;\psi_5 
= - \gamma\; m. \eqno(C.9)\] 
which are found to be nilpotent of order two (i.e. $s_{(a)b}^2 = 0$) and they are absolutely
anticommuting ($s_b \;s_{ab} + s_{ab}\; s_b = 0$) in nature as can be checked from the (anti-)BRST
symmetry transformations listed in (39) and (40).

We state, in passing, that the coupled (but equivalent) Lagrangians, their symmetries and their (anti-)BRST invariance
as well as the CF-type restrictions and their (anti-)BRST invariance, etc., for the massive spinning relativistic particle, can also be
captured within the framework of superfield formalism. This can be accomplished, in exactly same manner,
as we have done for the massless spinning relativistic particle in our present investigation.


\begin{thebibliography}{99}
\bibitem{TM1}    J. Thierry-Mieg, J. Math. Phys. {\bf 21}, 2834 (1980)
\bibitem{TM2}    J. Thierry-Mieg, Nuovo Cimento A {\bf 56}, 396 (1980)
\bibitem{QUHM}   M. Quiros, F. J. De Urries, J. Hoyos, M. L. Mazon, E. Rodrigues, \\
                 J. Math. Phys. {\bf 22}, 1767 (1981)
\bibitem{BT}     L. Bonora, M. Tonin, Phys. Lett. B {\bf 98}, 48 (1981) 
\bibitem{BT}     L. Bonora, P. Pasti, M. Tonin, Nuovo Cimento A {\bf 63}, 353 (1981)
\bibitem{DJ}     R. Delbourgo, P. D. Jarvis, J. Phys. A: Math. Gen {\bf 15}, 611 (1981)
\bibitem{DJT}    R. Delbourgo, P. D. Jarvis, G. Thompson, Phys. Lett. B {\bf 109}, 25 (1982)
\bibitem{RPM1}   See, e.g., R. P. Malik, Mod. Phys. Lett. A {\bf 20}, 1767 (2005)
\bibitem{RPM1}   See, e.g., R. P. Malik, Int. J. Mod. Phys. A {\bf 20}, 4899 (2005)
\bibitem{RPM1}   See, e.g., R. P. Malik, Eur. Phys. J. C {\bf 48}, 825 (2006) 
\bibitem{Nemsch} D. Namschansky, C. Preitschopf  and M. Weinstein, 
                 Ann. Phys. (N. Y.) {\bf 183}, 226 (1988)
\bibitem{dirac}  P. A. M. Dirac, {\it Lectures on Quantum Mechanics}, 
                 Belfer Graduate School of Science,\\
                 Yeshiva University Press, New York (1964)
\bibitem{sund}   K. Sundermeyer, {\it Constrained Dynamics: Lecture Notes in Physics}, 
                 vol. 169,\\ Springer-Verlag, Berlin (1982)         
\bibitem{RPM4}   R. P. Malik, Eur. Phys. J. C {\bf 45}, 513 (2006)
\bibitem{CF}     G. Curci, R. Ferrari, Phys. Lett. B {\bf 63}, 51 (1976)
\bibitem{BM1}    L. Bonora, R. P. Malik, J. Phys. A: Math. Thor. {\bf 43}, 375403 (2010) 
\bibitem{BM2}    L. Bonora, R. P. Malik, Phys. Lett. B {\bf 655}, 75 (2007) 
\bibitem{RPM5}   R. P. Malik, {\it etal.}, in preparation 
\end{thebibliography}
\end{document}